\documentclass[fleqn,usenatbib]{mnras}

\usepackage{newtxtext,newtxmath}
\usepackage[T1]{fontenc}

\DeclareRobustCommand{\VAN}[3]{#2}
\let\VANthebibliography\thebibliography
\def\thebibliography{\DeclareRobustCommand{\VAN}[3]{##3}\VANthebibliography}

\usepackage{graphicx}	% Including figure files
\usepackage{amsmath}	% Advanced maths commands
\title[DECam DDF II]{Deep drilling in the time domain with DECam II: characterizing the light curves of candidates in the extragalactic fields}

\author[M. L. Graham et al.]{
Melissa L. Graham,$^{1}$\thanks{E-mail: mlg3k@uw.edu}
Midori Rollins,$^{1}$
Robert A. Knop,$^{2}$
Suhail Dhawan,$^{3}$
Gloria Fonseca Alvarez,$^{4}$
\newauthor
Christopher A. Phillips,$^{1}$
Guy Nir,$^{2}$
Emily Ramey,$^{2}$
and Peter E. Nugent$^{2}$
\\
$^{1}$DIRAC Institute, Department of Astronomy, University of Washington, 3910 15th Avenue NE, Seattle, WA 98195, USA\\
$^{2}$E.O. Lawrence Berkeley National Laboratory, 1 Cyclotron Rd., Berkeley, CA, 94720 \\
$^{3}$Institute of Astronomy and Kavli Institute for Cosmology, University of Cambridge, Madingley Road, Cambridge CB3 0HA, UK\\
$^{4}$NSF NOIRLab, 950 N. Cherry Ave., Tucson, AZ 85719, USA
}

\date{Accepted XXX. Received YYY; in original form ZZZ}

\pubyear{2015}

\begin{document}
\label{firstpage}
\pagerange{\pageref{firstpage}--\pageref{lastpage}}
\maketitle

\begin{abstract}
In this second paper on the DECam deep drilling field (DDF) program we release 
2,020 optical $gri$-band light curves for transients and variables in the 
extragalactic COSMOS and ELAIS fields based on time series observations 
with a 3-day cadence from semester 2021A through 2023A.
In order to demonstrate the wide variety of time domain events detected 
by the program and encourage others to use the data set, 
we characterize the sample by presenting a brief analysis 
of the light curve parameters such as time span, amplitude, and peak brightness.
We also present preliminary light curve categorizations, and identify 
\emph{potential} stellar variables, active galactic nuclei,
tidal disruption events,
supernovae (such as Type Ia, Type IIP, superluminous, and
gravitationally lensed supernovae), and fast transients.
Where relevant, the number of identified transients is compared to the
predictions of the original proposal.
We also discuss the challenges of analyzing DDF data
in the context of the upcoming Vera C. Rubin
Observatory and its Legacy Survey of Space and Time, which will include DDFs.
Images from the DECam DDF program are available without proprietary
period and the light curves presented in this work are publicly available 
for analysis. 
\end{abstract}


%%%%%%%%%%%%%%%%% EXAMPLES %%%%%%%%%%%%%%%%%%%%%%%
% \begin{equation}
%     x=\frac{-b\pm\sqrt{b^2-4ac}}{2a}.
% 	\label{eq:quadratic}
% \end{equation}

% \begin{figure}
%     \includegraphics[width=\columnwidth]{example}
%     \caption{This is an example figure. Captions appear below each figure.
% 	Give enough detail for the reader to understand what they're looking at,
% 	but leave detailed discussion to the main body of the text.}
%     \label{fig:example_figure}
% \end{figure}

% \begin{table}
% 	\centering
% 	\caption{This is an example table. Captions appear above each table.
% 	Remember to define the quantities, symbols and units used.}
% 	\label{tab:example_table}
% 	\begin{tabular}{lccr} % four columns, alignment for each
% 		\hline
% 		A & B & C & D\\
% 		\hline
% 		1 & 2 & 3 & 4\\
% 		2 & 4 & 6 & 8\\
% 		3 & 5 & 7 & 9\\
% 		\hline
% 	\end{tabular}
% \end{table}

%%%%%%%%%%%%%%%%% BODY OF PAPER %%%%%%%%%%%%%%%%%%

\section{Introduction}\label{sec:intro}

The term "deep drilling" refers to an astronomical survey strategy in 
which a specific area of sky, typically a small field or the field-of-view of the 
telescope being used, is observed repeatedly with high frequency ("drilled") to 
detect time-variable sources and to build up depth by stacking the sequence of 
images.
This strategy is especially well-suited to explorations of the faint, 
fast, and high-redshift (high-$z$) universe of Galactic and extragalactic 
transients and variables.

Two of the most well known modern optical deep fields are
the Hubble Deep and Ultra Deep fields. 
Even without the "drilling" component and just two to five epochs of 
observations, high-$z$ supernovae\footnote{Granted, high-$z$ SNe can be 
found without the "drilling" component as they
last for weeks to months in the rest frame and benefit from an extended
visibility window thanks to cosmological time dilation.} (SNe) and 
active galactic nuclei (AGN) were discovered and analyzed
\citep[e.g.,][]{2003ApJ...589..693B, 2003ApJ...599..173S, 2004ApJ...613..200S, 2006AJ....131.1629S}.
Ground-based time-domain surveys also began obtaining 
deep observations of small-area fields with low cadence (weeks or more between
observations) to find high-redshift SNe
\citep[e.g.,][]{2004ApJ...602..571B, 2007MNRAS.382.1169P}.

The first optical multi-band time-domain survey that could be considered
to employ a "deep drilling" strategy was the Supernova Legacy Survey (SNLS), 
which was based on the four one-square-degree Deep fields of the
Canada France Hawaii Telescope (CFHT) Legacy Survey (CFHTLS; e.g.,
\citealt[][]{2006A&A...447...31A}).
At the time, the CFHTLS Deep field survey was referred to as a "rolling search"
to represent how the multi-band observations were repeated frequently
and continuously over the months when each of the four Deep fields was accessible. 
The rolling nature of the SNLS led to better characterizations for
Type Ia supernova (SN\,Ia) light curves, especially at early times,
which in turn led to more efficient use of the limited resources
for spectroscopic follow-up \citep{2006AJ....131..960S} and better constraints
on any cosmological evolution in the SN\,Ia light curves \citep{2006AJ....132.1707C}.

More recent examples of deep, well-cadenced time-domain optical imaging surveys
include the Pan-STARRS1 Medium Deep Survey \citep{2012ApJ...745...42T}
and the Dark Energy Survey Supernova program \citep[DES-SN;][]{2015AJ....150..172K}.
In the near future, the 10-year Legacy Survey of Space and Time (LSST)
executed by the new Vera C. Rubin Observatory will include at least five 
deep fields with a multi-band "deep drilling" strategy, along with its
wide-area time-domain coverage of the southern sky
\citep{2019ApJ...873..111I, 2022ApJS..258....1B}.
At the time of this publication the strategy had yet to be finalized, but
proposals included, e.g., an hour-long series of $\sim30$ second exposures for a 
single 9.6 deg$^2$ field which rotates through the $griz$ filters and
is repeated every 2-3 nights for a few months \citep{2019ApJ...873..111I},
or a combination of rolling surveys in ultradeep fields ($z<1.1$) with a
high cadence for a limited number of seasons and in deep fields ($z<0.7$)
with a three-night cadence every season \citep{2024arXiv240510781G}.
Regardless of the exact strategy adopted, the LSST deep drilling fields
will provide unprecedented access to 
time-domain astrophysics on timescales of minutes to days, and by
stacking the nightly sequences will provide access to the faint and high-$z$
variable and transient universe.

In order to prepare for the LSST era of time domain astronomy a new
deep drilling program with the Dark Energy Camera 
\citep[DECam;][]{2008SPIE.7014E..0ED} began in 2021.
The survey strategy, processing pipelines, and the first two 
semesters of data was presented in \citet{2023MNRAS.519.3881G}.
In this work, we present and make a preliminary analysis of the first 
five semesters of data for the two extragalactic deep drilling fields (DDFs)
and release 2,020 light curves in the $gri$ filters.
The main goal of this paper is to provide an overview of the data
available, and to encourage and enable public use.
In Section~\ref{sec:obs} we describe the observations and how objects
were detected and measured in difference images; the construction of
nightly-epoch light curves from the difference image photometry;
how we cross-matched to other transient- and static-sky catalogs of the 
extragalactic DDFs; and where these results are publicly accessible.
In Section~\ref{sec:sci} we make \emph{preliminary} classifications for
potential variables and transients, and illustrate the
difference-image photometry that is available by showing a variety
of representative light curves.
Full photometric classification with, e.g., template fits, machine
learning algorithms, or direct-image photometry is beyond the scope
of this paper and left for future work.
In Section~\ref{sec:conc} we conclude and describe future planned
work for the DECam DDF, which restarted observations in the 2024A semester.

\section{Observations}\label{sec:obs}

The overall design of the DECam DDF program and the real-time
data processing pipeline is described in detail in
\citep[hereafter refered to as Paper~I]{2023MNRAS.519.3881G}.
This program began in semester 2021A and ran through 2023A,
and at the time of writing was planned to start again in semester 2024A.
During all semesters, the DECam DDF was executed as one of
the programs of the DECam Alliance for Transients (DECAT):
a group of independent time-domain programs who pool their 
awarded time, are co-assigned nights, and share in the
observing duties to effectively run a queue service.
The proprietary period has been waived for this program's images.

As described in Paper~I, the DECam DDF covered two Galactic
and two extragalactic fields, COSMOS ($10\rm{h}$, $+2\deg$) and 
ELAIS ($0\rm{h}$, $-43\deg$).
These extragalactic fields have been previously covered by legacy
surveys across the electromagnetic spectrum. 
This work presents a preliminary characterization -- and releases 
for public analysis -- of difference-image light curves for
objects in the two extragalactic fields over five semesters 
from 2021A through 2023A (whereas Paper~I published only sources 
from 2021).

The DECam DDF observing strategy was to visit the COSMOS
and/or ELAIS field once every three nights while the field
was accessible (airmass $<1.5$, although higher-airmass visits
were done on rare occasions to fill scheduling gaps).
Each visit was composed of five sequences, each of which
obtained exposures at each of three (or two) adjacent pointings for 
COSMOS (or ELAIS): 80, 70, and 90 seconds in the 
$g$, $r$, and $i$ filters, respectively.
Each pointing thus received a total of 15 exposures per night.
The limiting magnitude was approximately $23.5$ mag for
each filter.

Processing pipelines to perform difference-imaging, source
detection and association, photometric calibration, real-bogus
classification and
alert production were adapted from the Dark Energy Survey
pipeline \citep{2019ApJ...881L...7G}.
The pipelines inputs, settings, and output photometric data products
are described in full in Section~3 of Paper~I.
In this work we adopt the same pipelines and data products,
and use the same terminology:
detections in individual difference-images are called
\emph{objects}, and objects that are within $1\arcsec$
of each other are associated into \emph{candidates}.

It is essential to understand that the pipelines currently
only consider \emph{positive} difference-image fluxes as
detected objects.
This is fine for transients which do not exist in the reference 
images (and thus always have a positive flux in the difference
image).
However, for astronomical phenomena like active galactic nuclei
or variable stars which can be present in the reference image
and can increase or decline in brightness, creating positive
or negative objects in the difference images, the candidate
light curves presented here will be incomplete (i.e., missing
the epochs when fainter than in the reference image).

\subsection{Data Access}\label{ssec:obs_dataaccess}

All images from the DECam DDF have no proprietary period
and can be retrieved from the NOIRLab Astro Data Archive\footnote{\url{https://noirlab.edu/public/projects/astrodataarchive/}}.
The full nightly-epoch light curves and the light curve summary parameters
described in Section~\ref{ssec:obs_nelc}, and the cross-match lists
in Section~\ref{ssec:obs_xmatch}, are all publicly available via
\texttt{GitHub}\footnote{Find all the data for this paper in the directory \texttt{transients\_science} within the repository at \url{https://github.com/MelissaGraham/decam_ddf_tools}}.

\subsection{Nightly-Epoch Light Curves}\label{ssec:obs_nelc}

Nightly-epoch light curves are created for candidates that
meet data-quality thresholds, as described in Paper~I.
To summarize here, the first step is to reject objects with 
real-bogus scores of $<0.1$ (79\% of all objects).
The second step is to reject candidates that are associated
with $<10$ objects (53\% of all candidates),
and then those which the mean real-bogus score of
all of their associated (unrejected) objects is $<0.4$
(52\% of the remaining candidates).
After these rejection steps, $2,020$ candidates remain.
A total of $1089$ candidates are in COSMOS, and $931$ are in ELAIS.

Nightly-epoch light curves are generated by grouping
objects by night and filter, and calculating their mean
time of observation, mean apparent difference-image magnitude, and 
mean real-bogus score.
The magnitude error for the nightly epoch is calculated by
adding in quadrature the mean of the object magnitude errors
and the standard deviation in the object magnitudes.
As of the end of the 2023A semester, the COSMOS field had 84 unique 
epochs and ELAIS had 97.
We also estimate a nightly-epoch limiting magnitude by combining
the 5$\sigma$ detection limits as measured in the individual
processed CCD images for each filter on the given night (Paper~I).

We use the term ``lonely epochs" to refer to nightly epochs 
for a given candidate
that have a mean real-bogus score of $<0.4$ and no other
nightly-epoch within 14 days that has a mean real-bogus score
of $>0.4$.
Nightly epochs that are flagged as ``lonely epochs" are not included
when calculating the light curve summary parameters (below)
because they are more likely to be spurious or unassociated with
the true astrophysical transient.
This use of ``lonely epoch" flags is new, and was not done in Paper~I.

We calculate four parameters to represent the
main characteristics of the nightly-epoch light curves:
the minimum difference-image magnitude (brightest epoch);
the amplitude (maximum minus minimum difference-image magnitude);
time span (days between the first and last difference-image 
detection); and number of non-lonely epochs.
These four parameters are determined for each filter individually 
($g$, $r$, and $i$) and over all filters.

\begin{figure*}
\includegraphics[width=\textwidth]{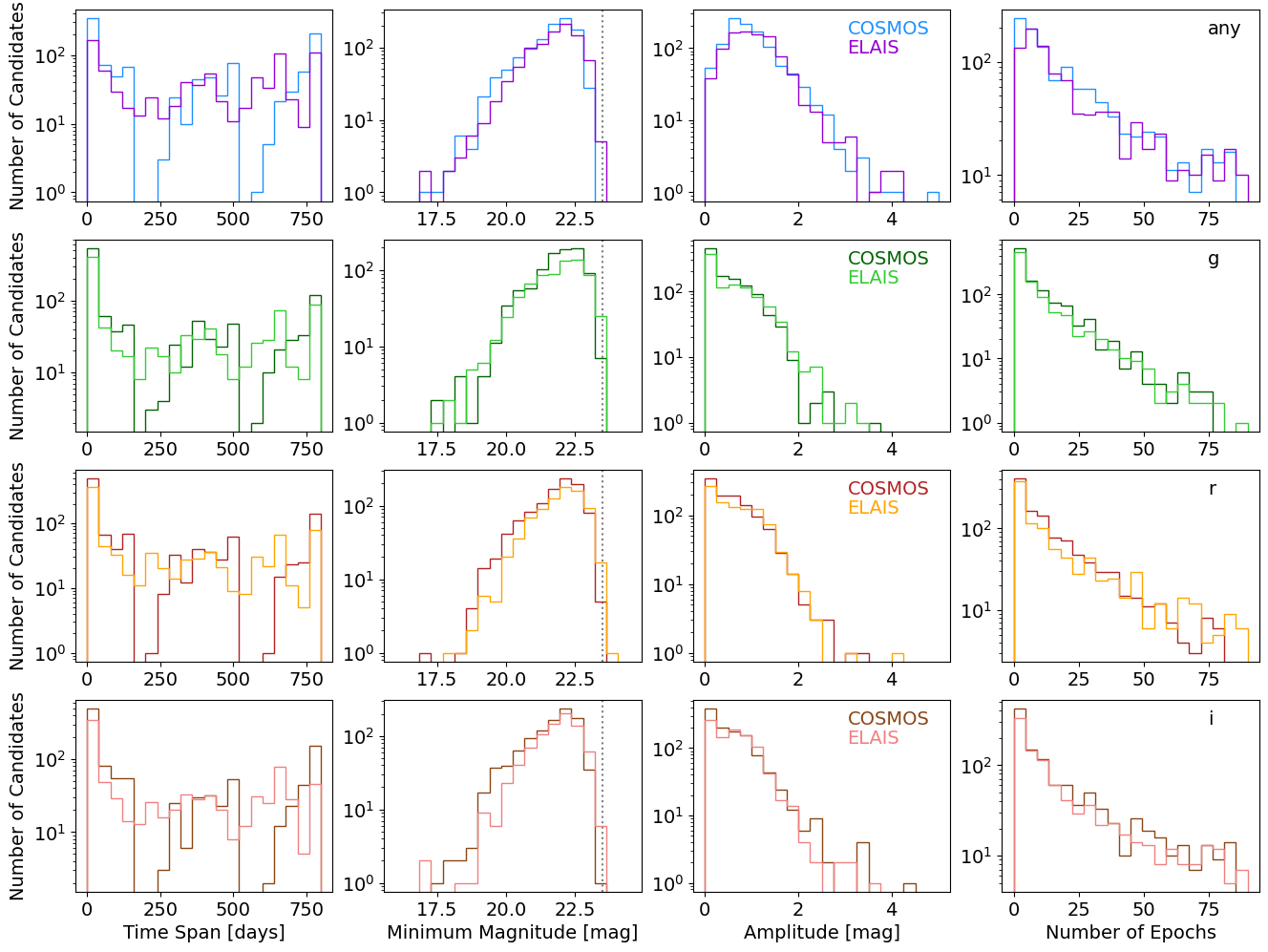}
\caption{Histograms for the nightly-epoch light curve parameters of all $2,020$ candidates that passed data-quality cuts.}
\label{fig:obs1}
\end{figure*}

\begin{figure*}
\includegraphics[width=\textwidth]{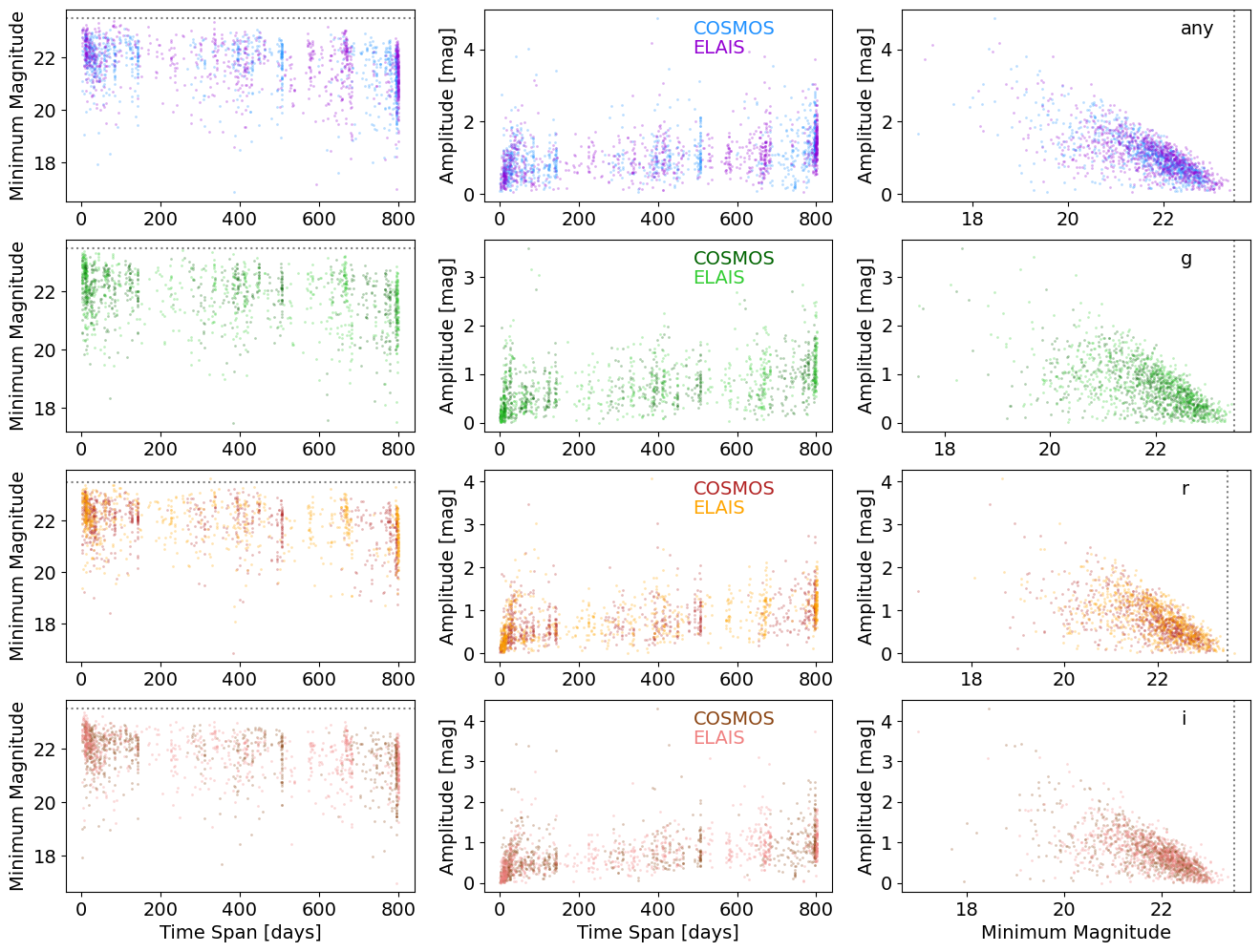}
\caption{Scatter plots for the nightly-epoch light curve parameters of all $2,020$ candidates that passed data-quality cuts.}
\label{fig:obs2}
\end{figure*}

In Figures~\ref{fig:obs1} and \ref{fig:obs2} we show the
histograms of the nightly-epoch light curve parameters,
and scatter plots of their relations with each other.
Some quantization in time span is seen in the COSMOS field,
which has shorter seasons from CTIO due to its more 
equatorial declination.
Dotted lines at the anticipated (theoretical) limiting magnitude
of the science (direct) images, $\sim$23.5 mag in each filter
(Section~\ref{sec:obs}; Paper~I), are drawn for plots which have
an axis of minimum magnitude.
This addition demonstrates that the faintest difference-image
detections align with the survey's limit.
It is these nightly-epoch light curves and their summary
parameters that are used as the starting point of our
analysis in Section~\ref{sec:sci}.

\subsection{Transient-Sky Counterparts}\label{ssec:obs_tns}

As there are many all-sky surveys detecting and reporting
transients, we use the Transient Name Server\footnote{https://www.wis-tns.org/}
to compile a list of classified transients that occurred
in the COSMOS or ELAIS field in the last three years.
We found five within the field boundaries, only one of which
happened while our program was observing: SN 2023dyl,
a Type Ia supernova, was detected as DC23klgla. 
However, since our program only detected five epochs in $i$-band
and one in $g$-band, this particular candidate did not
make it on to the list of potential SN\,Ia that is 
discussed in Section~\ref{ssec:sci_snia}.

\subsection{Static-Sky Counterparts}\label{ssec:obs_xmatch}

To identify static-sky counterparts (stars and host galaxies)
we cross-match our DECam DDF candidates to the Tractor 
Catalog from Data Release 10 (DR10) of the DESI Legacy Imaging 
Surveys\footnote{\url{https://www.legacysurvey.org}} 
\citep{2019AJ....157..168D}.
We perform the cross-matching in the JupyterLab environment at 
the Astro Data Lab at NSF's National Optical-Infrared 
Astronomy Research Laboratory (NOIRLab), using their 
{\tt queryClient} service. 

We start by simply counting the number of Tractor objects within a
radius of 72\arcsec\ ($0.02^{\circ}$) as a basic check that the 
catalog does cover the location of the candidate.
Since the locations of the DECam DDF were chosen, in part, based
on past DECam coverage the expectation was that all candidates'
coordinates would be covered by the Tractor catalog.
We confirmed this to be the case: there are 100 to 600 Tractor 
objects within a radius of $0.02^{\circ}$ of each of our candidates.

We then used the {\tt SkyCoord} class in the {\tt astropy} 
package to calculate the separation distance between each 
candidate and the Tractor objects within a radius of $0.02^{\circ}$. 
If the nearest Tractor object had a {\tt type} equal to PSF 
(i.e., a star), and it was within two times the size of the $r$-band 
point-spread function (PSF; column {\tt psfsize\_r}),
then the candidate was flagged as being cross-matched with a star.
If the nearest Tractor object was not a star but an extended object,
and the separation distance was less than five times the object's
half-light radius, then the candidate was flagged as being 
cross-matched to a potential host galaxy.

Of the 2,020 candidates, 1081 (53\%) were matched to a star,
802 (40\%) were matched to a potential host galaxy, 
and 137 (7\%) had no match.
Of the 802 candidates that were matched to a potential host galaxy,
619 (77\%) were flagged as potentially nuclear because the
separation distance was less than the PSF size.
These potentially nuclear candidates might be, e.g., transients in 
the cores of their host galaxy or active galactic nuclei (AGN).

Of the 1081 candidates cross-matched to a star, 477 (44\%) were 
also within 5 half-light radii of the nearest extended object.
Of the 802 candidates cross-matched to a galaxy, 93 (12\%) were
also within 2 times the PSF of the nearest star.
These candidates were flagged for being potentially mis-matched.

For all cross-matched objects we retained the following
columns from the DR10 Tractor catalog\footnote{For more information 
about the columns, see 
\url{https://www.legacysurvey.org/dr10/catalogs/}.}:
Right Ascension and Declination coordinates 
(columns {\tt ra} and {\tt dec}); 
Milky Way E(B-V) value \citep[\texttt{ebv},][]{1998ApJ...500..525S};
the Legacy Survey's unique identifier ({\tt ls\_id});
apparent magnitudes in filters $g$, $r$, and $i$ 
({\tt mag\_g}, {\tt mag\_r}, {\tt mag\_i}); 
the best-fit model {\tt type} (PSF; 
round exponential galaxy model, REX; 
de Vaucouleurs model, DEV; 
exponential model, EXP; 
or a Sersic model, SER); 
ellipticity parameters ({\tt shape\_e1} and {\tt shape\_e2});
the half-light radius ({\tt shape\_r});
the sersic index ({\tt sersic});
and the PSF size in filters $g$, $r$, and $i$
({\tt psfsize\_g}, {\tt psfsize\_r}, {\tt psfsize\_i}). 

For candidates cross-matched with a potential host galaxy, we use
a radius of 2\arcsec\ to cross-match the potential host to the objects 
in the DR9 Tractor and photometric redshift (photo-$z$) catalogs.
This step of cross-matching the DR10 and DR9 Tractor catalogs was 
necessary as the photo-$z$ were only available with DR9.
The use of 2\arcsec\ is appropriate in this case because we are
not cross-matching the \emph{transient} to the potential host galaxies
in the DR9 Tractor catalog, but cross-matching between galaxy 
central coordinates.
We find that 92\% of the potential host galaxies have a photometric
redshift, and that of these, 15\% also have a spectroscopic redshift.
From the DR9 Tractor and photo-$z$ catalogs we retain the columns
of mean and standard deviation of the photo-$z$ posterior distribution 
function (PDF; {\tt z\_phot\_mean} and {\tt z\_phot\_std}), as well
as the spectroscopic redshift ({\tt z\_spec}).

\begin{figure*}
\includegraphics[width=8cm]{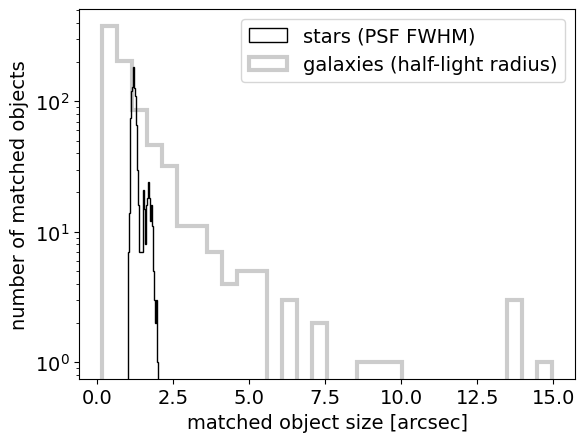}
\includegraphics[width=8cm]{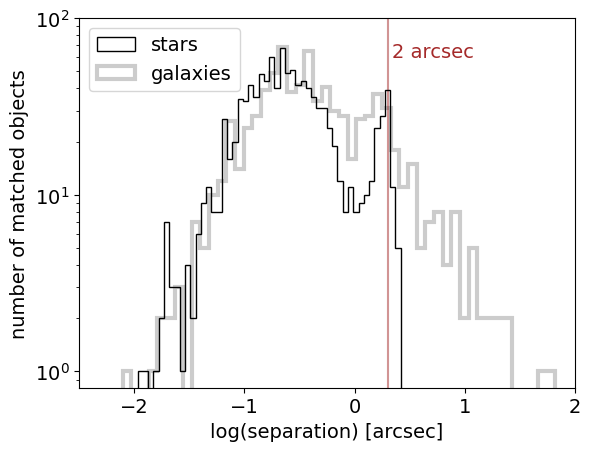}
\includegraphics[width=8cm]{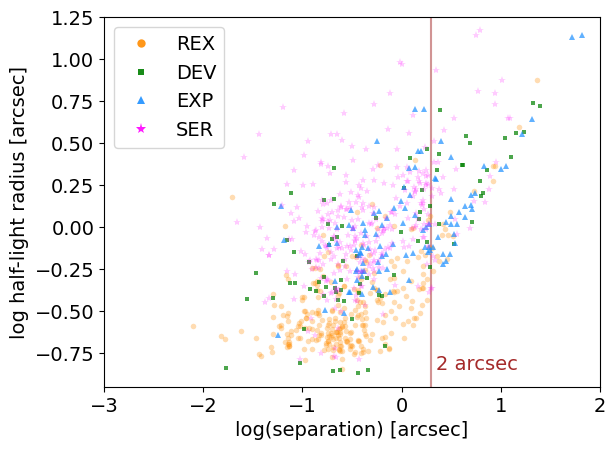}
\includegraphics[width=8cm]{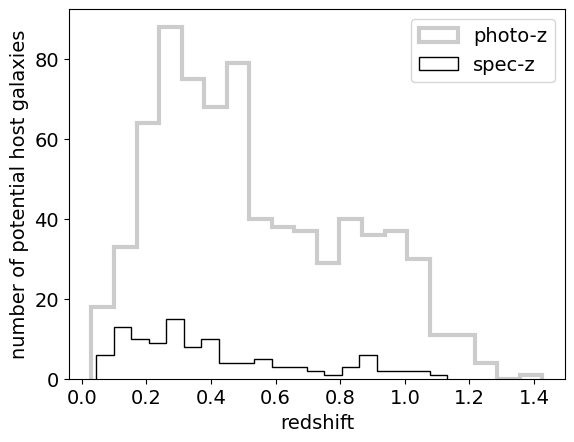}
\caption{Plots that characterize the Legacy Survey DR10 objects cross-matched
to the DECam DDF candidates.
The two upper panels include both matched stars and galaxies, while
the two bottom panels represent matched galaxies only.
At upper left, histograms of the matched object sizes, the PSF full-width
at half-maximum (FWHM) for stars and the half-light radius for galaxies.
At upper right, histograms of the separation distance  between candidate and
DR10 object, with an excess at 2\arcsec\, which is also the maximum image PSF size. 
At lower left, the half-light radius versus the separation for galaxies, with point
symbol and color based on the best-fit extended-source model.
At lower right, histograms of the photometric and spectroscopic redshifts from
DR9 for galaxies.}
\label{fig:obs3}
\end{figure*}

In Figure~\ref{fig:obs3} we provide some plots to illustrate the 
cross-matched population from the DR10 and DR9 Tractor catalogs.
At upper left, histograms of the matched object sizes for stars
show a range of PSF FWHM from about 1\arcsec\ to 2\arcsec. 
The distribution of PSF size is bimodal, and further investigation
(not shown) revealed this to come from only the ELAIS field.
Since the PSF sizes are actually the weighted average of the
PSF FWHM in a given band and not measured from the source 
directly, this bimodality represents the image quality 
(seeing) distribution of the survey in the ELAIS region.
The size distribution for potential host galaxies has many more 
small half-light radii ($<2\arcsec$) than larger extended objects, 
but this is expected as the candidates themselves are more likely 
to be intermediate to high redshift, or active AGN with bright cores. 
At upper right, the histogram of the separation distances for
cross-matched stars and galaxies exhibits a similar slope on the
small-separation end. 
There is an excess of candidates with a separation distance of
$\sim$2\arcsec\ from cross-matched stars and, to a lesser degree, galaxies.
% We speculate that this excess might indicate a systematic issue 
% (e.g., an offset) between the DECam DDF and DR10 astrometric solutions. 
% Since the cross-matching is used primarily for context and
% interpretation in Section~\ref{sec:sci}, and not, e.g., as 
% a prior or limit in any analysis, we don't explore this further in this work.
The bimodality in the separation distribution exists for both COSMOS and ELAIS, and is unrelated to the bimodality in the distribution of PSF size seen at upper left (i.e., there is no correlation between separation and PSF size for matched stars). To test if it was due to chance associations we generated 1000 random coordinates in the ELAIS and COSMOS fields, applied the same cross-match process, and found that the distribution of separations for coincidental matches peaked at $\sim$2\arcsec. The excess number of matched stars is $\sim$60, or about 3\% of the total number of candidates (although it does look outsized in the log-log plot). The static-sky matched objects are used for context in Section~\ref{sec:sci}, but none of our results depend strongly on it, so we find this amount of contamination acceptable.

In the lower-left panel of Figure~\ref{fig:obs3} we plot the 
half-light radius versus the separation for candidates with potential
host galaxies, with point styles representing the best-fit
extended-source models. 
This panel gives an overall impression of the
number of potential hosts with the different profiles, and shows
that candidates with more significant offsets ($>2\arcsec$) are 
typically in galaxies with de Vaucouleurs or exponential profiles.
This panel also shows a distinct group of matched objects that are
best-fit with Sérsic profiles, and have separations of 2\arcsec. 
% This is likely related to the excess of separations at 2\arcsec, 
% which is seen also for stars (upper-right panel).
The static-sky matched objects are used for context in
Section~\ref{sec:sci}, but none of our results depend strongly on it.
Finally, in the lower-right panel of Figure~\ref{fig:obs3} we show the
histograms of photometric and spectroscopic redshifts for potential
host galaxies.
Although the two distributions do have similar shapes, the
distribution of spec-$z$ is shifted towards lower redshifts and a
Kolmogorov-Smirnov test shows that they are not representative
of the same underlying distribution (i.e., are significantly different,
as expected).

\subsection{Spectroscopic Counterparts}\label{ssec:desi_xmatch}

We also cross-matched to the early data release of the
Dark Energy Spectroscopic Instrument, DESI \citep{2023arXiv230606308D}.
Of the 1,089 DECam DDF candidates in COSMOS, 597 (55\%) are 
matched to a DESI spectrum, with 131 of these spectra classified
as a galaxy, 432 as a QSO, and 15 as a star.
We find some discrepancies between the DESI spectroscopic
classifications and Legacy Survey photometric (morphological)
classifications for the static-sky counterparts.
For example, 41 of the cross-matched DESI galaxies were identified
as stars in the Legacy Survey.
This simply represents the known challenge
of photometric classifications in deep surveys.

\begin{figure}
\includegraphics[width=8cm]{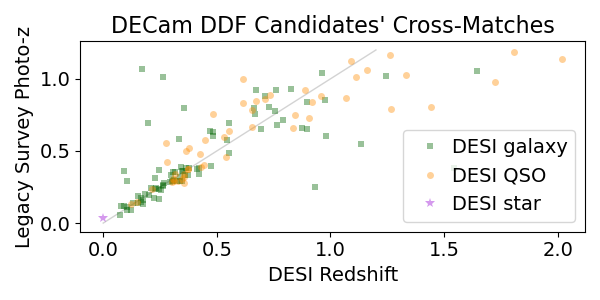}
\includegraphics[width=8cm]{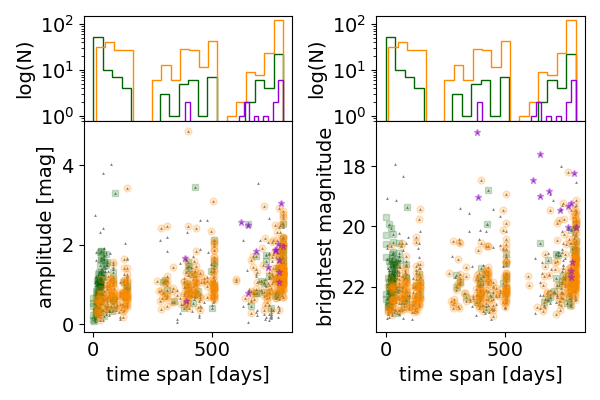}
\caption{Top: the Legacy Survey photometric redshifts versus the DESI 
redshifts for static-sky objects cross-matched to the DECam DDF candidates. 
Symbols and point color represent the DESI classifications as in the legend. 
Bottom: scatter plots of the DECam DDF light curve parameters for all 
candidates in COSMOS with multi-night detections (small triangles), and for 
candidates cross-matched to DESI objects.
Histograms demonstrate how the time span distributions for galaxies
(green) and QSO (orange) are similar.}
\label{fig:obs4}
\end{figure}

To further characterize the DESI and Legacy Survey static-sky
counterparts, we present a comparison of the photometric
and spectroscopic redshifts in the top panel of Figure~\ref{fig:obs4}.
Interestingly, it seems that some of our DECam DDF candidates are
at higher redshifts ($z>1.5$) than indicated by the photometric
redshifts alone; these will be explored further in future work.
There is also one object classified as a star by DESI, but
which was classified as a galaxy in the Legacy Survey's 
Tractor catalog, albeit with a very low photometric redshift.
This object is a bright blue point sources, and most likely a star.

To further characterize the DECam DDF candidates with DESI matches,
in the bottom panel of Figure~\ref{fig:obs4} we show scatter
plots of the light curve parameters (as in Figure~\ref{fig:obs2})
for all candidates in COSMOS with multi-night detections.
We can see that the DECam DDF candidates which are matched
to DESI galaxies generally have shorter time spans;
this makes sense as these candidates are likely transients in
host galaxies which were targeted by DESI.
On the other hand, DECam DDF candidates which are matched
to DESI stars and QSOs have longer time spans and,
for the longer-duration events, larger amplitudes and brighter
difference-image minimum magnitudes -- as expected for 
variables like stars and QSOs.

Examples of DECam DDF light curves for hosted transients,
QSOs (AGN), and variable stars are shown in Section~\ref{sec:sci}.
We leave for future work the analysis of the time-domain
counterparts of the DESI spectra, and here just comment that
all of these data are publicly available.

\section{Candidate Characterization}\label{sec:sci}

Photometric classification for variable and transient
phenomena is a fast-developing field in astronomy, and 
will be even more essential as future wide-area sky surveys
like the LSST which detect millions of time-domain events -- 
far more than can be followed-up spectroscopically.
While we are working towards installing and running 
published classifiers on the DECam DDF candidates\footnote{E.g., the \texttt{ParSNIP} classifier as presented in \citet{2021AJ....162..275B}.}, this paper is focused on simple ways to use the 
light curve parameters discussed in Section~\ref{ssec:obs_nelc}
to identify and characterize \emph{potential} candidate populations.
It is our purpose and hope that this initial analysis will
inspire others to take and use the candidate light curves,
which are publicly available (Section~\ref{ssec:obs_dataaccess}).

\subsection{Potential Stellar Variables}\label{ssec:sci_stars}

As mentioned in Section~\ref{sec:obs}, the light curves produced
by the processing pipelines and used for this 
work only include epochs in which the candidate had a 
\emph{positive} difference-image flux.
This precludes most typical analyses of variable stars,
such as periodigrams.
Instead, for this work we provide a preliminary
characterization of variable stars in the extragalactic
DECam DDFs.

Periodic variables would appear in our data set as
candidates with a long time span, but with detections
in only a fraction of the epochs (i.e., when the star
is brighter than it is in the reference image).
Periodic variables would exhibit a difference-image
light curve of alternating detections and non-detections
on a timescale of days for short-period variables but
weeks to months for longer-period variables.
Aperiodic variables that are always brighter in the
DECam DDF images than in the reference image might 
be detected in every epoch, and have varying or 
similar difference-image magnitudes over the years.

\begin{figure}
\centering
\includegraphics[width=\columnwidth]{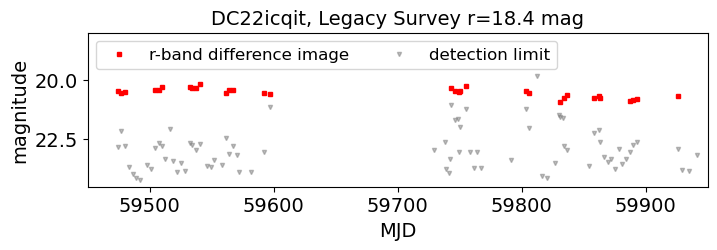}
\includegraphics[width=\columnwidth]{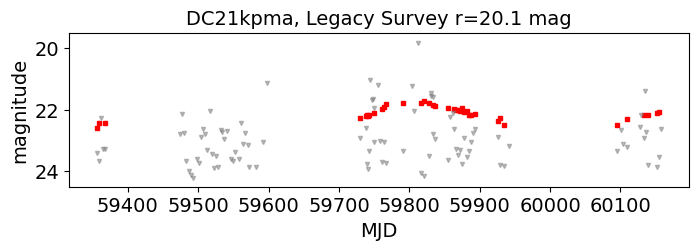}
\includegraphics[width=\columnwidth]{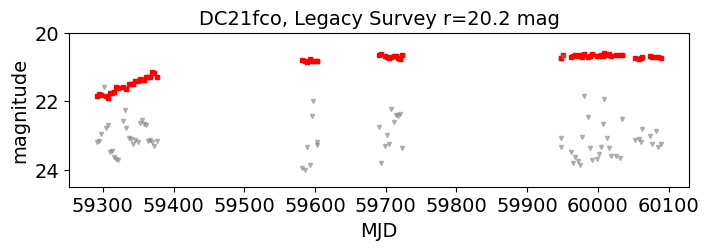}
\caption{The $r$-band difference-image light curves (for positive difference-image flux detections only; red squares) for a few selected potential stellar variables in the DECam DDF extragalactic fields. Small grey symbols represent the $r$-band nightly-epoch limiting magnitude estimates.}
\label{fig:sci_stars_lperiodic}
\end{figure}

In Figure~\ref{fig:sci_stars_lperiodic} we show 
example difference-image $r$-band light curves for
three of the 1081 candidates which were matched to stars in 
the Legacy Survey (Section~\ref{ssec:obs_xmatch}).
To identify these three representative examples,
we visually reviewed the 34 which exhibited
$r$-band light curve time spans $\geq 750$ days, 
amplitudes $\geq 0.5$ mag, minimum magnitudes $\leq 22$ mag,
and detected in $\geq 70$ epochs.
In the top panel, DC22icqit appears and disappears
as a positive flux source in the difference images
on a timescale of days, and could be a short-period variable.
In the middle panel, DC21kpma appears to have a slowly
varying positive difference-image flux over the 
2022A and 2022B semesters (MJD $\sim$59750 to $\sim$59950).
However, it was not detected during the 2021B semester 
(MJD $\sim$59480 to $\sim$59600), presumably due to 
it being fainter than in the reference image.
In the bottom panel we show DC21fco as an example of a
potential aperiodic variable star which was always
brighter in the DECam DDF images than in the reference image.

\subsection{Potential Active Galactic Nuclei}\label{ssec:sci_agn}

Active Galactic Nuclei (AGN) are known to be variable sources.
Optical variability can be used to identify AGN, as well as estimating
the mass of the central black hole and accretion disk sizes through
techniques like reverberation mapping
\citep[see][for a review of reverberation mapping]{Cackett21}.
Variability in the inner disk propagates outwards, with the same variability
observed at longer wavelengths at a later time.
This makes large scale photometric surveys useful to study the structure of
accretion disks for a large number of AGN. 

As mentioned in Section~\ref{sec:obs} and discussed with respect to 
variable stars above, the light curves produced
by the processing pipelines and used for this 
work only include epochs in which the candidate had a 
\emph{positive} difference-image flux. 
While this precludes a physical analysis of AGN, it is still possible to
make a preliminary characterization of potential AGN candidates in the
extragalactic DECam DDFs for demonstration purposes.
\citet{2024arXiv240206052Z} present AGN analysis from 
their High-quality Extragalactic Legacy-field Monitoring (HELM) program,
which includes COSMOS and ELAIS DECam DDF images.

\subsubsection{Cross-matches to AGN in SIMBAD catalogs}

As the COSMOS and ELAIS deep drilling fields are legacy fields
with a long history of astronomical observations, many 
of the AGN have already been detected through various techniques
(e.g., spectroscopy, photometric variability), in different wavelengths (e.g., X-rays).
As presented in \citet{2023AAS...24110709R},
we retrieve a heterogeneous sample of AGN in COSMOS or ELAIS from 
a variety of catalogs available via SIMBAD 
\citep{2000A&AS..143....9W}:
the COSMOS2015 galaxies catalog \citep{2016ApJS..224...24L},
the Advanced Camera for Surveys General Catalog 
\citep[SCSGC;][]{2012ApJS..200....9G},
the ESO Spitzer Imaging extragalactic survey 
\citep[ESIS;][]{2006A&A...451..881B},
and the Spitzer Wide-Area Infrared Extragalactic Survey 
\citep[SWIRE;][]{2013ApJS..208...24L}.

We find that 466 of the 2,020 candidates are matched to an 
already-known AGN.
Most are classified as quasi-stellar objects (QSOs; 325) or 
AGN (119), and a few are Seyfert galaxies (20).
The catalog we retrieved from SIMBAD is both very
heterogeneous and contains many duplicate sources,
and we consider it beyond the scope of this work to 
draw inferences about, e.g., the fraction of AGN which were
identified by spectroscopy or X-rays emission that are
optically variable (or vice versa).

\subsubsection{Potential AGN light curves}

Variability in AGN can be described as stochastic, and has typical
timescales of months and years, with average variability in the order
of 10-20\% \citep{VanderBerk04}.
To find candidates that are potential AGN, we apply constraints on the
nightly-epoch light curves based on general characteristics of AGN variability.
We expect they would show up for the duration of the survey, disappearing for
periods where the field is not visible, or when the AGN brightness is fainter
or not significantly different than the template.
Therefore we set a minimum time span of two years, a minimum difference-image
magnitude of 23 mag, a minimum amplitude of 0.2 mag, and a minimum number
of 40 epochs in any filter ($g$, $r$, or $i$).
Since AGN emission is nuclear, we add the constraint that if the candidate
is matched with a galaxy based on the Tractor catalog, then it should also be
in the core of the galaxy.
There are 254 candidates that meet the restrictions on their
photometric variability from the DECam DDF, and 131 are cross-matched to
already-known AGN.
Adding the restriction that these conditions be met on all three filters
reduces the number of candidates to 54 and 28 respectively.

\begin{figure}
\centering
\includegraphics[width=\columnwidth]{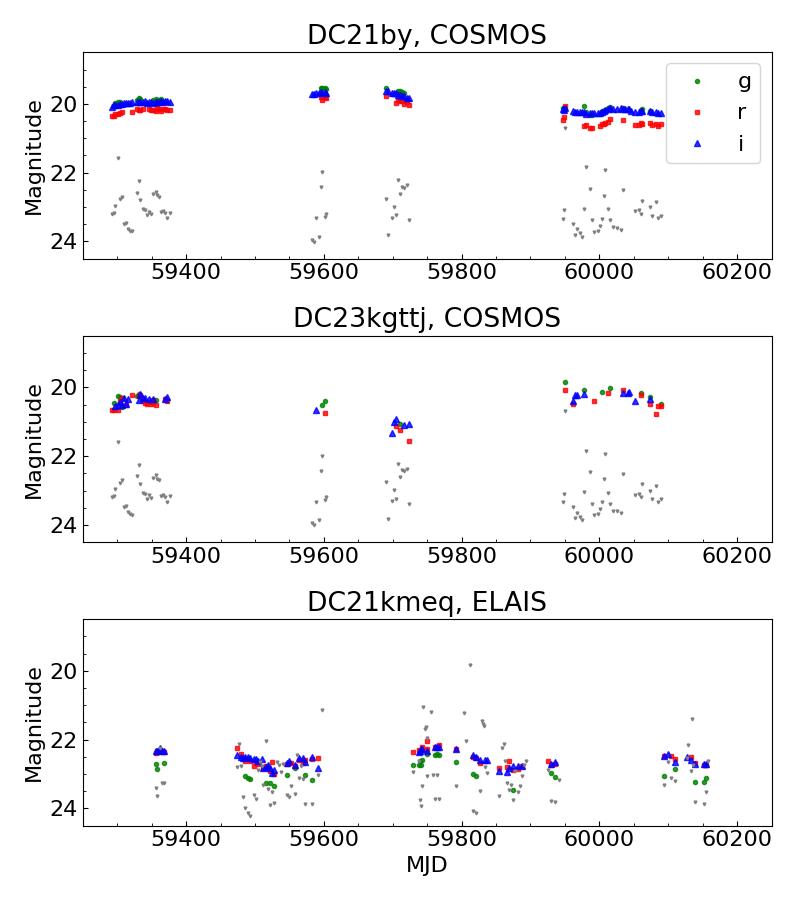}
\caption{Difference image $g$, $r$,$i$ photometry vs modified Julian Date (MJD) 
light curves for three candidates.  DC21by is a known QSO, and DC23kgttj and 
DC21kmeq are examples of potential AGN in the COSMOS and ELAIS DDF. 
Small grey symbols represent the $r$-band nightly-epoch limiting magnitude estimates.}
\label{fig:AGN_char_LC}
\end{figure}

\begin{figure}
\centering
\includegraphics[width=\columnwidth]{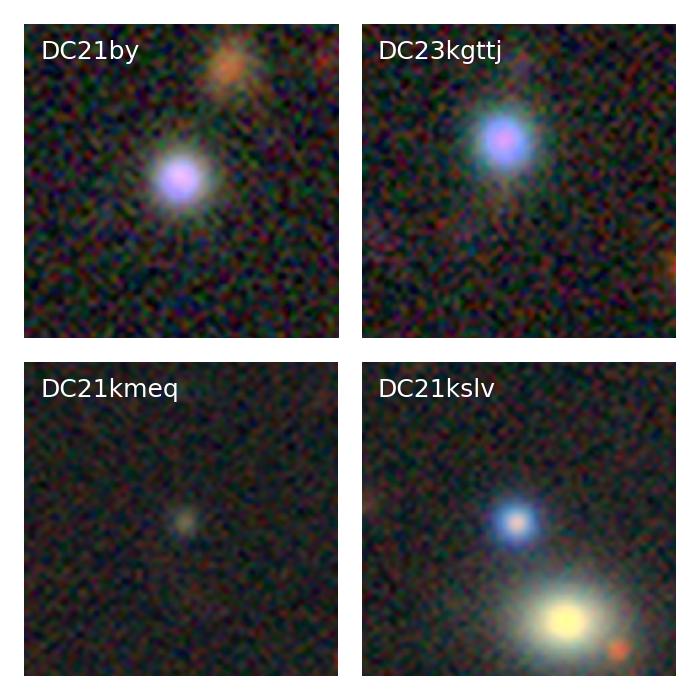}
\caption{Color images from the Legacy Survey DR9 at the location of four 
candidates selected for their AGN-like variability with light curves shown 
in Figures \ref{fig:AGN_char_LC} and \ref{fig:DC21kslv_LC}.}
\label{fig:AGN_DR9_images}
\end{figure}

To demonstrate what potential AGN light curves look like in the DECam DDF's
difference-image photometry, Figure~\ref{fig:AGN_char_LC} shows the light
curves of three candidates that display stochastic, long term variability
in any of the $g$, $r$, $i$ bands, and are matched to a galaxy core.
In the top panel, DC21by is a known QSO in the COSMOS field.
Gaps in the data points are partly due to observing gaps (roughly corresponding
to semesters 2021B and 2022B), and partly due to the candidate being fainter or
of similar brightness to the template image.
Two additional light curves were chosen to show in Figure~\ref{fig:AGN_char_LC}
due to their similarity with known QSO DC21by.
In the middle panel, DC23kgttj shows a similar light curve to DC21by, but is
not matched to a known AGN.
It is briefly not detected in the $i$-band around MJD 60000, when it probably
had a brightness similar to or fainter than the template image.
In the bottom panel, DC21kmeq is another example of a potential AGN that was
not matched to a known AGN in the ELAIS field.
Gaps in the data points roughly coincide with semesters 2021A, 2022A and 2023A.
Around MJD 59800, it is possible it was too faint for detection.
In Figure \ref{fig:AGN_DR9_images}, we show color images from the Legacy Survey
DR9 at the location of these three candidates, all of which are point-like
and suggest that DC21kmeq and DC23kgttj are QSO like DC21by.

\begin{figure}
\centering
\includegraphics[width=\columnwidth]{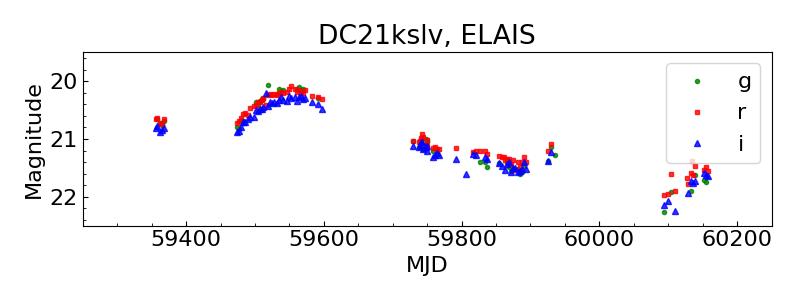}
\caption{Difference image $g-$, $r-$, and $i-$band photometry versus modified Julian Date (MJD) lightcurves for variable nuclear candidate DC21kslv.}
\label{fig:DC21kslv_LC}
\end{figure}

In our visual review of the potentially AGN-like candidates that were
cross-matched to the center of their potential host, the light
curve of DC21kslv stood out as different from the potential AGN
shown in Figure~\ref{fig:AGN_char_LC}.
The light curve of DC21kslv, shown in Figure \ref{fig:DC21kslv_LC},
exhibits a smooth rise over $\sim$70 days, with the peak occurring at a
magnitude of 20.06 (in the $g$-band), around MJD $\sim$59520, and then
slowly declines -- at first glance, appearing almost transient-like.
However, the duration of the rise is longer than for supernovae and
longer than seen for tidal disruption events (TDE) like, e.g., PS1-10jh \citep{2012Natur.485..217G}.
Although at first glance DC21kslv looks special, given its early detections 
around MJD 59350 days, its late time rise at 60100 days, and its
stellar-like appearance in Figure \ref{fig:AGN_DR9_images},
we suspect it is probably just a QSO.

The takeaway message from this preliminary analysis is
that the DECam DDF does provide rich time-domain information
for AGN for future analyses, but that the applications of
difference-image photometry alone for AGN are limited.

\subsection{Potential Type IIP Supernovae (SN\,IIP)}\label{ssec:sci_ccsn}

Type II supernovae (SN\,II) are caused by the core collapse of massive stars.
SN\,IIP light curves are heterogeneous in their peak brightness, but
typically exhibit a $\sim$100-day "plateau" phase of slow decline 
after peak brightness \citet{2012ApJ...756L..30A},
which is unique among supernova types.
This plateau occurs when the hydrogen envelope surrounding the progenitor 
star is retained before core collapse \citep{Smartt_2004}.
In our original proposal for the DECam DDF, we had estimated finding
$\sim$5 SN\,IIP per semester based on volumetric explosion rates
(i.e., with no loss due detection or classification efficiencies).

To find potential SN\,IIP among our candidates, we first identify 243
candidates with $r$-band light curve time spans between 50 and 150 days.
We calculate the $r$-band light curve slope in magnitudes per day for
observations between 15 and 80 days after the first detection.
This would be the "plateau" phase for any candidates that are SN\,IIP.
We then restrict to only those candidates with slopes of less than
$0.03$ magnitudes per day, over at least 40 days, and with at least
7 epochs (nights detected). 
This is still a very generous range of parameter space, as a full SN\,IIP
light curve would decline by $<$3 magnitudes over 100 days and, with 
the DDF's cadence, likely be detected on $>$7 nights.

\begin{figure}
    \centering
    \includegraphics[width=\columnwidth]{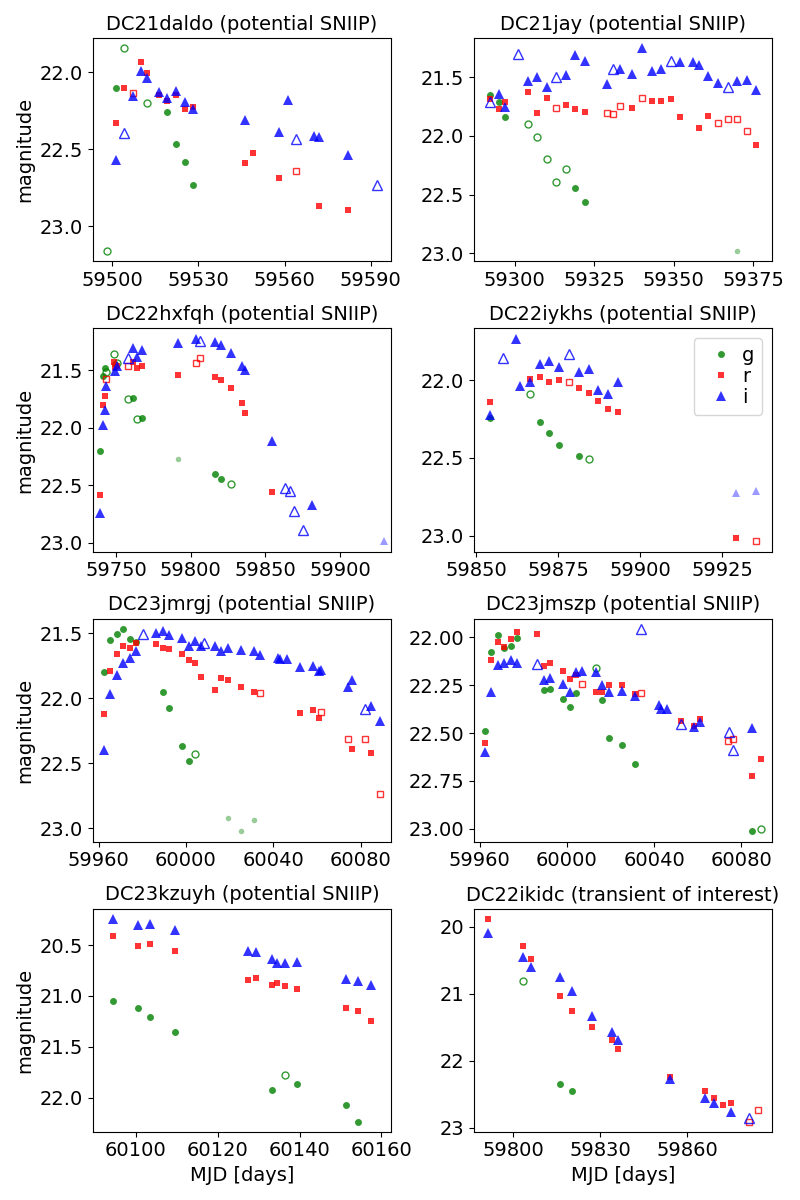}
    \caption{The DECam DDF multi-band light curves for seven potential SN\,IIP identified with simple restrictions on the light curve parameters, and (at lower right) one transient of interest (DC22ikidc) which also made the initial cuts, but is not SN\,IIP-like. Open symbols represent a nightly-epoch photometry point with a mean real-bogus score of $<$0.4 (closed symbols represent $\geq$0.4, and transparent closed symbols represent "lonely epochs" as described in Section~\ref{ssec:obs_nelc}.}
    \label{fig:sci_sniip_lcs}
\end{figure}

These simple, generous restrictions result in 42 candidates with
a slow decline over at least a few weeks post-peak for visual review.
Of the 42, most appear to be active galactic nuclei in that their
light curves rise and decline non-monotonically, but 7 do meet the description
of a potential SN\,IIP in their time span and slow decline.
The light curves for these seven are shown in Figure~\ref{fig:sci_sniip_lcs},
along with one additional candidate of interest which was identified
from among the 42 as a potential tidal disruption event
(Section~\ref{sssec:sci_tde_DC22ikidc}).
In this set, DC22hxfqh is an oddity with a long-duration rise in the 
$i$-band over a $\sim$90-day "plateau" phase.

The utility of a deep-drilling style program is evident from the 
shorter-timescale features that are on display in the light curves
in Figure~\ref{fig:sci_sniip_lcs}.
For example, DC21daldo, DC21jay and DC23jmszp exhibit light curve "bumps" that are 
reminiscent of Type IIn supernovae, in which the ejecta material
interacts with circumstellar material and can cause re-brightenings
as was seen for, for example, SN\,2009ip \citep[e.g.,][]{2013MNRAS.430.1801M,2013ApJ...767....1P,2014ApJ...787..163G}.
We note that although the "bump" in the light curve for DC23jmszp at around
40 days past peak is a similar timescale to the first bump in the light
curve of SN\,2009ip, the two events are actually not overall that similar
because SN\,2009ip declined by $\sim$2 magnitudes on that time scale,
and DC23jmszp does not (and neither do DC21daldo or DC21jay).
Another example of the benefit of the deep-drilling style program
is that for most of these seven potential SN\,IIP,
the $g$-band declines much more rapidly but thanks to the 2-4 cadence
of the DECam DDF, is still well-sampled in this dataset.

Finding seven potential SN\,IIP over five semesters is well off the
prediction from the original proposal of $\sim$5 \emph{per semester},
but notice in Figure~\ref{fig:sci_sniip_lcs} that our rough restrictions
have yielded only the brightest events.
These seven all reach peaks brighter than 22 mag, whereas our predicted
yields were based on the detection limits of $\sim$23.5 mag.
Longer duration transients like SN\,IIP are one of the event types whose detection 
efficiency with this DDF program would benefit significantly from
difference imaging with nightly stacked images, which remains 
work in progress (Section~\ref{sec:conc}).

\begin{figure}
    \centering
    \includegraphics[width=\columnwidth]{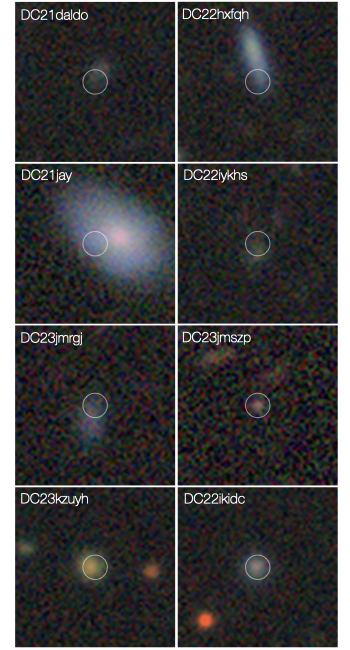}
    \caption{The locations of the eight candidates whose light curves are shown in Figure~\ref{fig:sci_sniip_lcs} in the Legacy Survey DR9 color image (oriented north up, east left). These stamps are all $\sim$14\arcsec x14\arcsec and centered on the candidates' coordinates, for which a light circle is drawn to guide the eye.}
    \label{fig:sci_sniip_hosts}
\end{figure}

\begin{table}
\centering
\begin{tabular}{lcccccc}
\hline
 & & \multicolumn{4}{c}{------\textbf{Tractor Catalog Match}------} & \\
\textbf{Name} & \textbf{Peak $m_r$} & \textbf{Type} & \textbf{$m_r$} & \textbf{$z_{\rm phot}$} & \textbf{$M_r$} \\
\hline
DC21daldo & 21.93 & SER & 22.38 & 0.37$\pm$0.15 & -19.6 \\
DC21jay   & 21.62 & EXP & 22.31 & 0.29$\pm$0.38 & -19.3 \\
DC22hxfqh & 21.39 & SER & 20.18 & 0.12$\pm$0.06 & -17.3 \\
DC22iykhs & 21.98 & EXP & 21.94 & 0.48$\pm$0.11 & -20.2 \\
DC23jmrgj & 21.57 & SER & 21.12 & 0.25$\pm$0.08 & -18.9 \\
DC23jmszp & 21.97 & PSF & 22.88 & N/A           &  N/A  \\
DC23kzuyh & 20.41 & REX & 21.04 & 0.47$\pm$0.04 & -21.7 \\
\hline
\end{tabular}
\caption{For the seven candidates identified as potential SN\,IIP based 
on their light curves, the peak $r$-band magnitude is shown in column 2. 
Characterization parameters for the matched object from the Tractor catalogs 
in column 3 to 5 are the best-fit model type, the static-sky $r$-band 
magnitude and the photometric redshift (as described in Section~\ref{ssec:obs_xmatch}). 
In column 6 we show the estimated peak absolute brightness for the candidate 
based on the photometric redshift, assuming a flat cosmology.}
\label{tab:sci_sniip}
\end{table}

\subsubsection{Cross-matches to the Tractor catalog}\label{sssec:sci_ccsn_xmatch}

Of the seven candidates identified, all of them are matched to an object in
the Tractor catalog, as shown in Table~\ref{tab:sci_sniip}.
Six of these objects are classified as galaxies, but the object at the 
location of DC23jmszp is classified as a star.
Color image stamps for all seven (plus the candidate of interest DC22ikidc)
are shown in Figure~\ref{fig:sci_sniip_hosts}.
They generally show the bluer spiral (or irregular) galaxies with ongoing
star formation that is typical for the hosts of core-collapse supernovae.
For the six candidates matched to galaxies with photometric redshifts
(none of the matches have spectroscopic redshifts),
the inferred peak $r$-band absolute magnitudes range from -17 up to almost -22.
For SN\,II, peak intrinsic brightnesses from -14 to -19 mag
can be considered as within the range of expectation
\citep[e.g.,][]{2011MNRAS.412.1441L,2014ApJ...786...67A,2015ApJ...799..208S, 2019A&A...631A...8H}. 
However, for example DC23kzuyh with an estimated
peak of -21.7 mag would be more appropriate for a superluminous
supernova (SLSN; \citealt{2007ApJ...668L..99Q,2019ARA&A..57..305G}).

\subsection{Candidate DC22ikidc: A Tidal Disruption Event?}\label{sssec:sci_tde_DC22ikidc}

\begin{figure}
\centering
\includegraphics[width=\columnwidth]{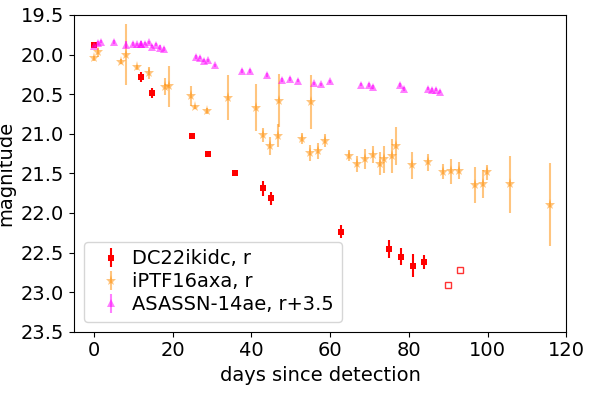}
\caption{The DECam DDF $r$-band light curve of candidate DC22ikidc (red squares) compared to he $r$-band light curves for TDEs ASASSN-14ae (pink triangles) and iPTF16axa (orange stars).}
\label{fig:sci_sniip_dc22ikidc}
\end{figure}

As described in Section~\ref{ssec:sci_ccsn},
DC22ikidc was identified as a potential SN\,IIP based on broad cuts to the
light curve parameters.
However, its light curve stands out from the other SN candidates' in 
Figure~\ref{fig:sci_sniip_lcs} because its long-term near-linear (in magnitudes)
decline is more reminiscent of a tidal disruption event (TDE).
In Figure~\ref{fig:sci_sniip_dc22ikidc} we compare the DECam DDF's $r$-band
light curve of DC22ikidc to the $r$-band light curves for two TDEs from the
literature with well-sampled declines: ASASSN-14ae \citep{2014MNRAS.445.3263H}
and iPTF16axa \citep{2017ApJ...842...29H}.
The two comparison light curves have been shifted in date, and ASASSN-14ae
has been shifted in magnitude, to align with DC22ikidc.
Unfortunately there is no way to constrain the event start date for DC22ikidc,
as the nearest epoch with a limiting magnitude was 22 days before detection 
(it was $r$$\sim$23.7 mag).

Although DC22ikidc does decline more rapidly than the two TDEs shown in
Figure~\ref{fig:sci_sniip_dc22ikidc}, considerable diversity is emerging
among TDEs \citep[e.g.,][]{2020SSRv..216..124V,2023A&A...673A..95C}.
Without spectra or observations across the electromagnetic spectrum
it will be hard to confirm the true nature of DC22ikidc.
However, it was located right in the core of its host galaxy, 
as shown in the lower-right panel of Figure~\ref{fig:sci_sniip_hosts}.
In the Tractor catalog, the host galaxy has a type of REX,
an apparent $r$-band magnitude of 21.35 mag, and a photometric redshift
of 0.22$\pm$0.10.
This photometric redshift gives DC22ikidc an intrinsic
magnitude of about -20$\pm$1 mag (the error is based on the estimated error
in the photometric redshift), which is typical for TDEs
\citep[e.g.,][]{2020SSRv..216..124V}.

\subsection{Potential Type Ia Supernovae (SN\,Ia)}\label{ssec:sci_snia}

In our original proposal for the DECam DDF, we had estimated
finding (per semester) $\sim$20 SN\,Ia within a redshift of $z<0.5$,
and one to two $z<0.25$ SN\,Ia within four days of explosion.
Although these estimates were for the originally-proposed
2-day cadence, did not include detection or classification
efficiencies, and were for anticipated processing that used deeper 
nightly image stacks for the difference-image detections
(which is still in development),
they are still in the ballpark for what we might expect from
the implemented DECam DDF survey.
As we did not have a program to spectroscopically follow-up 
DECam DDF targets, and we are leaving photometric classification
(using software based on e.g., machine learning or template-fitting)
to future work, we instead develop an informal 
process to identify \emph{potential} SN\,Ia among the candidates.

Since SN\,Ia light curves are homogenous, events at lower
redshifts reach brighter apparent magnitudes and are 
detectable for longer (in surveys with a constant limiting
magnitude, like the DECam DDF).
Using template light curves from \citet{2002PASP..114..803N} we
derive an approximate linear relationship (for the DDF's limiting
magnitudes) between light curve amplitude $A$ and time span $D$
for filter $f$, such that SN\,Ia have $A_f > m_f D_f + b_f$
where the slope values are $m_g = 0.050$ and $m_r = m_i = 0.037$
mag per day and the intercepts are
$b_g = b_i = -1.5$ and $b_r = -1.125$ mag.
Light curve amplitude and time span are two of the light curve
parameters we had already calculated for all candidates
(Section~\ref{ssec:obs_nelc}), so these approximate relations
allow us to use these parameters to identify potential SN\,Ia.
We take as potential SN\,Ia all candidates which meet the
following conditions in at least two of the three filters:
light curve time span is $\geq 10$ and $\leq 200$ days;
amplitude $\geq 0.5$ mag; 
number of nights detected $\geq 5$; 
and $A_f > m_f D_f + b_f$ for filter $f$.
This yields 100 potential SN\,Ia out of all our DECam DDF candidates.

To all 100 potential SN\,Ia we apply the \texttt{sncosmo}
light curve fitting code \citep{barbary_kyle_2023_8393360}.
We chose \texttt{sncosmo} because it is python-based and works
off-the-shelf with optical light curves.
With \texttt{sncosmo} we use the \texttt{SALT2} model
\citep{2007A&A...466...11G}.
The idea was to use \texttt{sncosmo} to fit the light curves
as if they were SN\,Ia, and then use the fit parameters to 
make a probabilistic assessment of a candidates' likelihood 
of being a SN\,Ia.
Before running \texttt{sncosmo} we visually review the light
curves of all 90 potential SN\,Ia candidates and 
flag those which are not well-sampled enough to expect a decent
light curve fit with \texttt{sncosmo}.
Candidates are flagged if their light curve is missing a filter,
missing the rise or the decline in all three filters,
the decline is sampled with $<$3 epochs or $\lesssim$15 days
in two or more filters,
or if the light curve does not approximately 
monotonically rise and fall like a transient (i.e., if it 
fluctuates more like an AGN).
Of the 100 potential SN\,Ia, 34 have unflagged light curves.

As an input to \texttt{sncosmo} we estimate the redshift 
of the candidate by assuming that it
is a SN\,Ia, that its brightest observed magnitude is its
peak magnitude, and that the intrinsic peak brightness of
SN\,Ia is -19.3 magnitudes.
We then pass upper and lower boundaries on redshift of $\pm0.15$
from this estimate to \texttt{sncosmo}.
If the lower boundary is $<0.05$, which is the lower limit of
\texttt{sncosmo}, then boundaries of $0.05$ to $0.35$ are 
passed instead (i.e., the same range width).
We acknowledge that starting off with an assumption that the
candidate is a SN\,Ia,
and passing a redshift based on that assumption, 
means that \texttt{sncosmo} is more likely to fit a SN\,Ia 
light curve, even if the event was another type of supernova.

\begin{figure}
\centering
\includegraphics[width=\columnwidth]{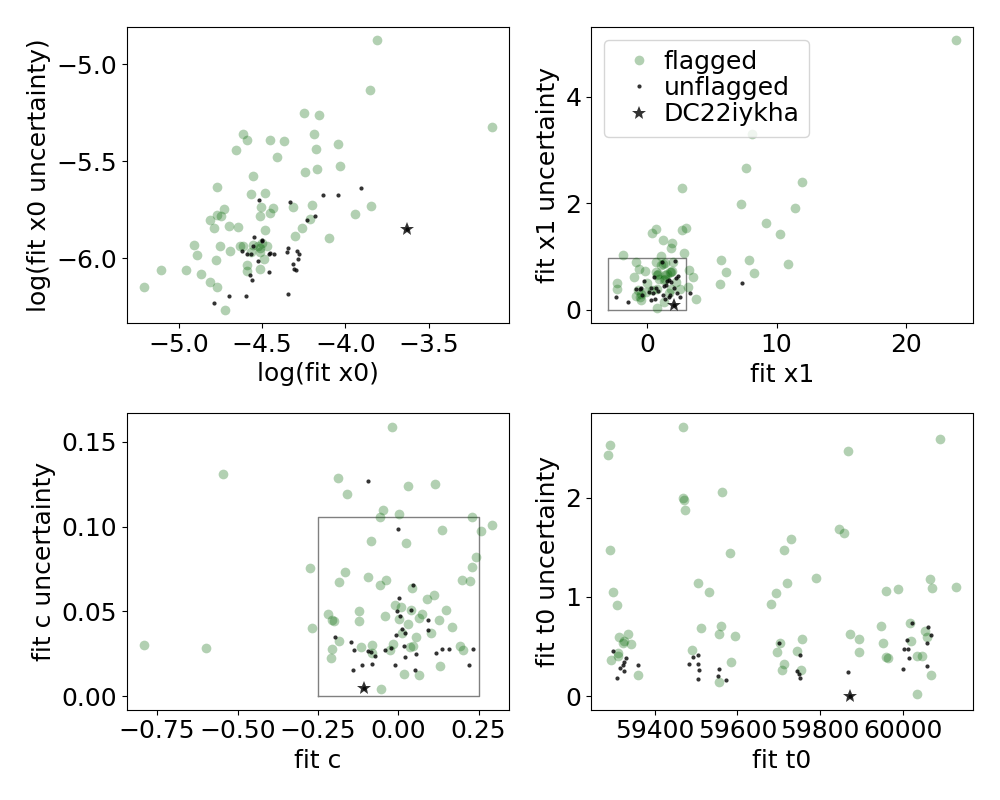}
\caption{The \texttt{sncosmo} best-fit \texttt{SALT2} model
light curve parameters of $x_0$ (related to peak brightness), 
$x_1$ (related to light curve decline rate),
$c$ (related to color at peak brightness), 
and $t_0$ (time of peak brightness), for the 66 
flagged (green circles) and 34 unflagged (black dots)
potential SN\,Ia candidate light curves.
The black star is the special case of candidate DC22iykha.
Grey boxes represent the typical ranges of the $x_1$ and $c$
parameters for SN\,Ia (x-axis limits) and the uncertainties
for unflagged light curves (y-axis limits).
There are 62 candidates with parameters in both boxes;
31 each of flagged and unflagged candidates.}
\label{fig:sci_snia_lcpars}
\end{figure}

In Figure~\ref{fig:sci_snia_lcpars} we show scatter plots of 
the SALT2 light curve parameters fit by \texttt{sncosmo} for
the 66 flagged and 34 unflagged potential SN\,Ia candidates.
These SALT2 parameters are $x_0$ (related to peak brightness), 
$x_1$ (related to light curve decline rate),
$c$ (related to color at peak brightness), 
and $t_0$ (time of peak brightness).
From these plots we can see that the idea to use the fit values
and uncertainties to assess whether a candidate was a SN\,Ia
was a bit naive.
As shown in Figure~\ref{fig:sci_snia_lcpars}, the flagged
candidates (green circles) do have higher uncertainties and
a broader ranges of fit parameters returned.
In some cases these are beyond the ranges for true SN\,Ia,
which are marked by the boxes
in the upper-right and lower-left panels: 
$-3 < x_1 < 3$ and $-0.25 < c < 0.25$
\citep{2009ApJ...700.1097H, 2023ApJ...953...35G}.
However, in general there is significant overlap between the
unflagged and flagged candidates.
In other words, even a poorly-sampled candidate light curve
can still return SN\,Ia-like fit parameters.
In total, there are 62 candidates in the boxes in the 
$x_1$ and $c$ panels of Figure~\ref{fig:sci_snia_lcpars},
31 each of flagged and unflagged candidates.

The main takeaway of this analysis is that overall, the
DECam DDF caught 62 candidates in 2 years that are decently fit 
as a SN\,Ia using the \texttt{SALT2} model and \texttt{sncosmo}.
The original estimate of $\sim$20 per semester ($\sim$100 over
five semesters) was off by about a factor of 2.
This is not surprising, given that this work is using detections
in single images instead of nightly coadds, and that the cadence
was lower than initially anticipated. 
The good news, however, is that if proper photometric
classification is applied to this now-public data set,
SN\,Ia science relating to rates or host galaxies should be
possible.
All are welcome to use this list of 100 potential SN\,Ia as a 
starting point for further work.

\subsubsection{Candidate DC22iykha}

\begin{figure}
\centering
\includegraphics[width=\columnwidth]{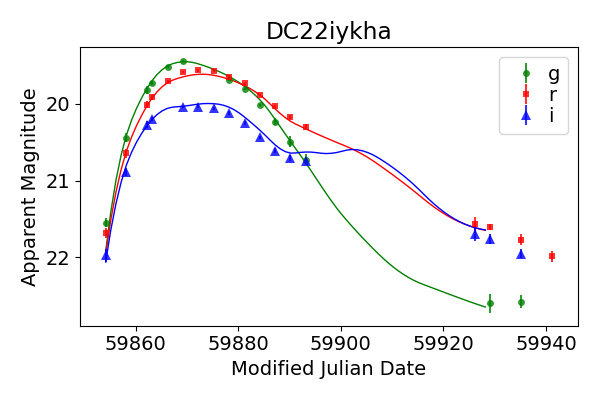}
\includegraphics[width=7cm]{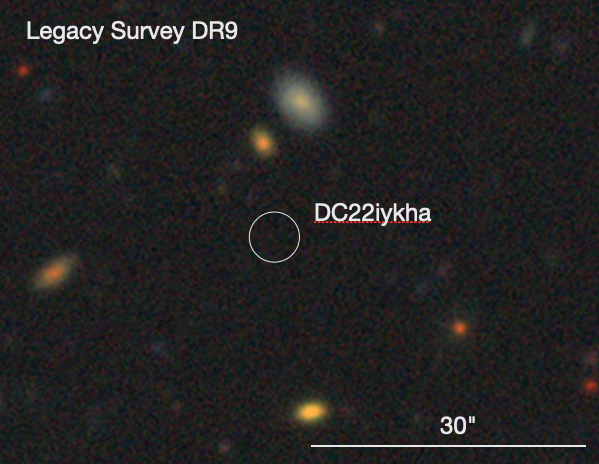}
\caption{Top: The DECam DDF multi-band light curve of candidate
DC22iykha (points), and the best fit \texttt{SALT2} model 
SN\,Ia light curve by \texttt{sncosmo} (lines).
Bottom: The location of candidate DC22iykha 
in the Legacy Survey DR9 color image (oriented north up, east left)
reveals it to be potentially hostless.}
\label{fig:sci_snia_DC22iykha}
\end{figure}

One particular candidate of interest, DC22iykha, is a
bright outlier among the potential SN\,Ia, and is marked
with a black star symbol in Figure~\ref{fig:sci_snia_lcpars}.
DC22iykha has a bright high-fidelity light curve, as
shown in the top panel of Figure~\ref{fig:sci_snia_DC22iykha}.
Although at first glance the \texttt{sncosmo} fit does seem
appropriate, the error bars on the photometry are small and the 
systematic offset in the $r$- and $i$-bands is significant.
Although the $i$-band does give a hint of the onset of the
$i$-band bump that is common in SN\,Ia light curves, the
apparent $r-i$ color is bluer than a typical SN\,Ia
(which is closer to 0).
We checked if the host galaxy or environment held any additional
information for this candidate by reviewing the Legacy Survey
DR9 color cutout\footnote{\url{https://www.legacysurvey.org/}}
shown in the bottom panel of
Figure~\ref{fig:sci_snia_DC22iykha},
and found that this candidate is potentially hostless.
The cross-matching with the Tractor catalog described in
Section~\ref{ssec:obs_xmatch} finds this candidate to be offset
by 1.7\arcsec\ from a galaxy that has a photometric
redshift of 1.07 (but an uncertainty of 0.59), 
which is not visible in the color image shown in
Figure~\ref{fig:sci_snia_DC22iykha}.
If it exists, it is more likely a background galaxy that is
unassociated with candidate DC22iykha.
A cross-match with the Transient Name
Server\footnote{\url{ttps://www.wis-tns.org/}}
reveals that this event was discovered and reported by
the Asteroid Terrestrial-impact Last Alert System
\citep[ATLAS;][]{2018PASP..130f4505T}
on 2022-10-20 and was assigned the designation AT\,2022yfq
\citep{2022TNSTR3057....1T}.

\subsubsection{Cross-matches to the Tractor catalog}\label{sssec:sci_snia_hosts}

Of the 100 candidates identified as potential SN\,Ia,
only three are not matched to a Tractor catalog object,
and for 84 the best match is a galaxy.
For the remaining 13 the best match is a star, but for ten
of these there is also a galaxy that is near enough to be
considered a potential host.
Considering only the 34 unflagged candidate light curves of
potential SN\,Ia, only one is not matched (DC23jsner; but see
Section~\ref{ssec:sci_unhosted}) and 29 are matched to a galaxy.
Four are matched to a star, but for three there is also
a galaxy near enough to be considered a potential host;
one is another potentially hostless transient (DC21jfcb).
This work has not used any information about potential host
galaxies (or stellar counterparts) to identify potential
SN\,Ia, nor used the potential host photometric redshifts
as a prior in any of the \texttt{sncosmo} fits.
However, we mention the cross-matches here to illustrate the
fact that there is likely useful potential-host information
that could and should be incorporated into future photometric
classification efforts and SN\,Ia analyses.

\subsubsection{Potential SN\,Ia Caught Early}\label{sssec:sci_snia_early}

One of the science goals of the DECam DDF was to find transients,
and in particular SN\,Ia, within a few days of explosion.
In our original proposal, we estimated that the DECam DDF could
catch $\sim1$ SN\,Ia at $z\lesssim0.25$ within four days of 
explosion per semester, but again this was based on the originally
proposed two-day cadence.
Multi-band photometry of SN\,Ia at early stages can exhibit
signatures of the binary companion star, of circumstellar
material, and/or of the mixing of radioactive material into the
outer layers of the exploding white dwarf star
(as discussed in, e.g., 
\citealt{2010ApJ...708.1025K, 2016ApJ...826...96P, 2017ApJ...845L..11H, 2019ApJ...870L...1D}).
In this work we make a preliminary investigation of how many
potential SN\,Ia were detected at early times, whether any
obvious deviation from a template light curve is evident.

\begin{figure}
\centering
\includegraphics[width=\columnwidth]{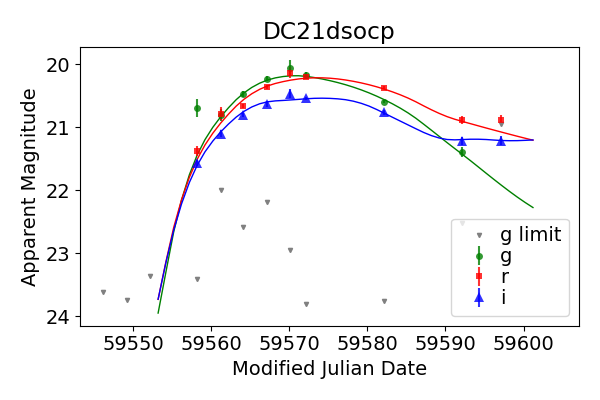}
\caption{The DECam DDF multi-band light curve of candidate
DC21dsocp (points), and the best fit \texttt{SALT2} model 
SN\,Ia light curve by \texttt{sncosmo} (lines).
Grey inverted triangles represent the $g$-band nightly-epoch limiting magnitude estimates.}
\label{fig:sci_snia_DC21dsocp}
\end{figure}

SN\,Ia light curves are known to have rise times (the time between
explosion and peak brightness) of around 19 days, plus or minus
a few days \citep{1999AJ....118.2675R,2015MNRAS.446.3895F}.
Cosmological time dilation will also increase the rise times
by a few days at $z=0.2$ (i.e., by a factor of $1+z$).
Of the 34 unflagged potential SN\,Ia identified, 
we find that 17 were first detected
between 14 and 28 days before the date of best-fit peak brightness
from \texttt{sncosmo}.
Of the 17, four have a fit redshift and peak brightness that
suggest they are $z\lesssim0.25$ events, which does
approximately match with the $\sim1$ per semester that
was originally predicted.
One of these four, DC21dsocp, exhibits a potential deviation
from its best-fit template light curve in its earliest
$g$-band detection, as shown in
Figure~\ref{fig:sci_snia_DC21dsocp}.
This potential deviation is later in time, as in
closer to peak brightness, than for SN\,Ia for
which early-time deviations have been previously detected
\citep[e.g.,][]{2017ApJ...845L..11H, 2019ApJ...870L...1D}.
Without spectroscopy or a fuller light curve, it is 
unclear even whether DC21dsocp is a SN\,Ia; perhaps
the $g$-band excess is a shock breakout or other phenomenon
for a core collapse SN.
It is beyond the scope of this work to say any more than
such photometric anomalies are detectable in this data set.

The DECam DDF observation blocks were designed such that,
during a DECAT night, they did not need to be done 
consecutively (i.e., all observations obtained within an hour).
Having the option to spread the DDF blocks out was intended to
both help with scheduling multiple programs and to potentially make
the intra-night photometry more useful by raising the potential
of detecting intra-night rising or declining events.
However, in practice, all DDF blocks were usually obtained within
an hour.
Of the 17 candidates detected $>14$ days before peak, only five
had two or more detections in the individual difference images
in at least two bands in their first night of detection.
In all cases these detections were all obtained within an hour
of each other and no intra-night rise is discernable (as might
be detectable if they happened to be caught within hours
of explosion).

\subsection{Potential Fast Transients}\label{ssec:sci_ft} 

Fast transients are short time-scale transients whose brightness
rises and falls within days to weeks.
Types of fast transients include fast blue optical transients
\citep{2014ApJ...794...23D}, and the optical counterparts of short gamma-ray
bursts gravitational wave events such as kilonovae (KN).
In the original proposal for the DECam DDF, we estimated finding
up to 1 rapidly evolving transient per semester.

To identify potential fast transients among the DECam DDF
candidates, we first identify candidates with a light curve time
span parameter less than 30 days in \emph{every} filter, and which
in at least one filter was detected at least three nights;
reached a peak brightness $\leq$22 mag; had an amplitude parameter
of $\geq$0.5 mag; and had a rise time of $\leq$10 days.
This results in 21 candidates for visual review.
After rejecting candidates that have multiple peaks, a slow
decline, or only a rise, we are left with four candidates.
None of the 21 potential fast transients we have identified declined as
quickly as the KN associated with gravitational wave event
GW170817, which declined by about 4 magnitudes
in 10 days in $g$- and $i$-band \citep{2017ApJ...848L..17C}.
The three-day cadence of the DECam DDF is a bit too long 
to confidently identify KN. 

\begin{figure}
    \centering
    \includegraphics[width=\columnwidth]{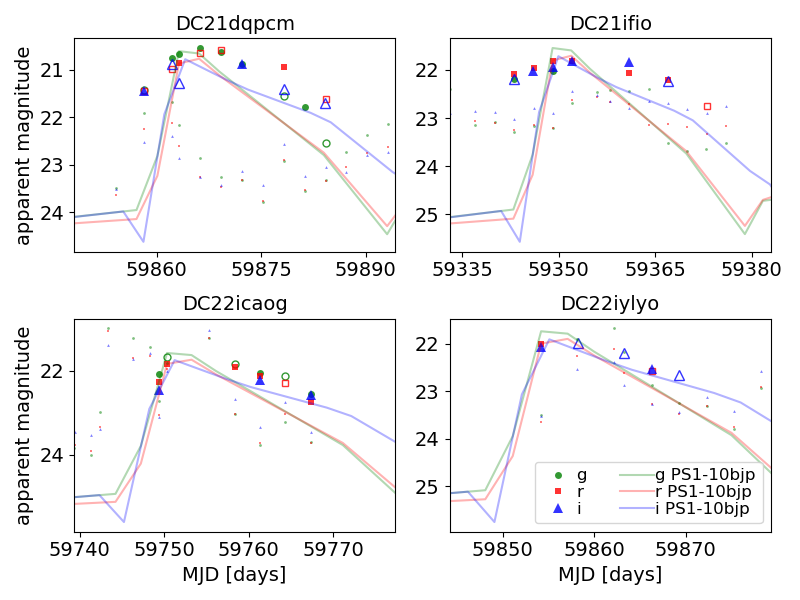}
    \caption{The DECam DDF multi-band light curves for four candidates identified with light curve parameter cuts as potential fast transients. Large symbols have the same meaning as in Figure~\ref{fig:sci_sniip_lcs} and small symbols represent the $r$-band nightly-epoch limiting magnitude estimates. The solid lines represent the photometry of PS1-10bjp from \citet{2014ApJ...794...23D}, shifted to approximately match in the $r$-band.}
    \label{fig:sci_ft_lcs}
\end{figure}

In Figure~\ref{fig:sci_ft_lcs} we show the DECam DDF light curves
for these four candidates, along with the light curves of fast
transient PS1-10bjp from \citet{2014ApJ...794...23D} for comparison.
PS1-10bjp was chosen for being the best observed object of the sample, and
it has been shifted to approximately match each DDF candidate in the $r$-band.
We can see that the two candidates in the top panels of
Figure~\ref{fig:sci_ft_lcs}, DC21dqpcm and DC21ifio,
are probably not fast transients after all, as
they rise and decline more slowly than PS1-10bjp.
While the candidates in the two lower panels, DC22icaog and DC22iylyo,
appear to match PS1-10bjp fairly well, DC22icaog lacks detection limits
to confirm the fast decline and DC22iylyo lacks pre-detection limits
to confirm a fast rise.

We additionally check the individual difference-image photometry from the
first night that DC22icaog was detected, in case
the DDF blocks were spread out sufficiently to identify an
intra-night rise.
As was the case for SN\,Ia (as described at the end of
Section~\ref{sssec:sci_snia_early}) no intra-night evolution in
the magnitude of DC22icaog was detectable.

\begin{figure}
    \centering
    \includegraphics[width=\columnwidth]{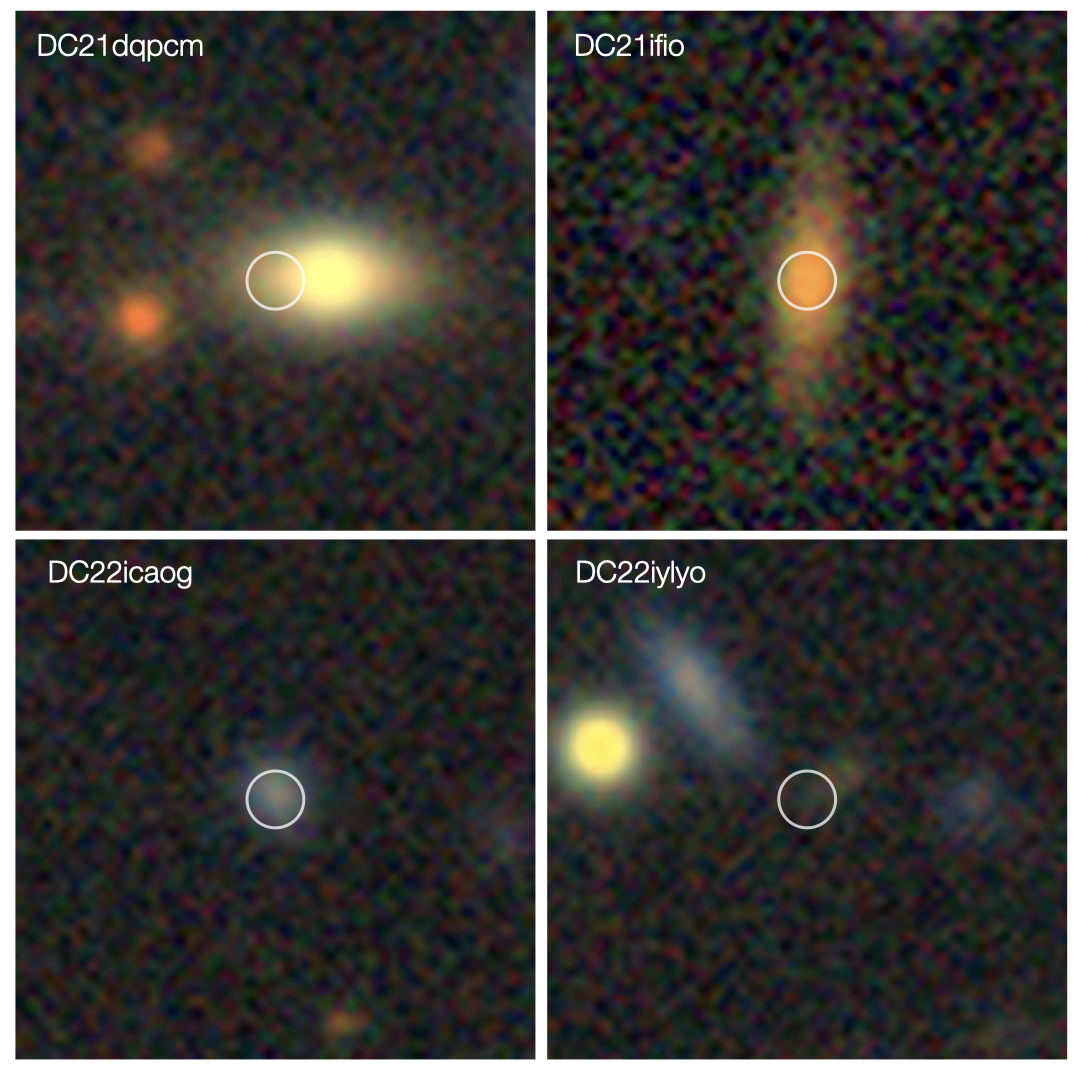}
    \caption{The locations of the four candidates whose light curves are shown 
    in Figure~\ref{fig:sci_ft_lcs} in the Legacy Survey DR9 color image, 
    in the same manner as Figure~\ref{fig:sci_sniip_hosts}.}
    \label{fig:sci_ft_hosts}
\end{figure}

All four of these candidates are cross-matched to galaxies in the
Tractor catalog, as shown in Figure~\ref{fig:sci_ft_hosts}.
These galaxies have photometric redshifts which, under the assumption
of a flat cosmology, result in estimated peak absolute magnitudes in
the $r$-band of -19.19 mag for DC21dqpcm, -19.43 mag for DC21ifio, 
-18.67 mag for DC22icaog, and -20.19 mag for DC22iylyo.
These estimates are consistent with the range of -16.5 to -20.0 mag for fast
transients from \citet{2014ApJ...794...23D}.
However, given the poor constraints before the first (and after the last)
detections for each of these candidates, ultimately none can be confirmed as
potential fast transients.

\subsection{Potential Lensed or Superluminous Supernovae}\label{ssec:sci_lens}

We search the sample for potential rare and bright objects like strongly gravitationally 
lensed supernovae. 
Due to magnification from lensing, they are expected to be brighter than the nominal 
absolute magnitude for their transient class at the source redshift. 
For objects like Type Ia supernovae, which have a small dispersion in their peak 
brightness, this is a powerful method to discover such object \citep{Goldstein2017}. 
We note that based on previous predictions \citep[e.g.,][]{Goldstein2019} we also 
expect to find many lensed core collapse SNe (CCSNe), however, since their peak 
luminosity has a large intrinsic dispersion, we require a different selection 
method to identify them in the data. 
Therefore, for this analysis, we only focus on the lensed SNe~Ia, but will expand 
the search to include all types of lensed SNe in future work.
The method based on the magnification has been used in the real-time discovery of 
the two galaxy scale lensed Type Ia supernovae discovered in wide field surveys, 
namely iPTF16geu \citet{Goobar2017} and SN~Zwicky \citep{Goobar2023}. 
Within this search method, we start with the subsample that has a spectroscopic 
redshift, so that the distance modulus to the source -- and hence the absolute 
magnitude -- can be accurately measured. 
We also analyse the subsample of objects that have a photometric redshift with 
an accuracy better than $\sigma(z) = 0.02$ corresponding to a distance modulus 
error $< 0.1$ mag at typical redshifts $\sim 0.5 - 1$ where we expect the SNe 
to be observed \citep{Arendse2023}.

\begin{figure}
    \centering
    \includegraphics[width=.48\textwidth]{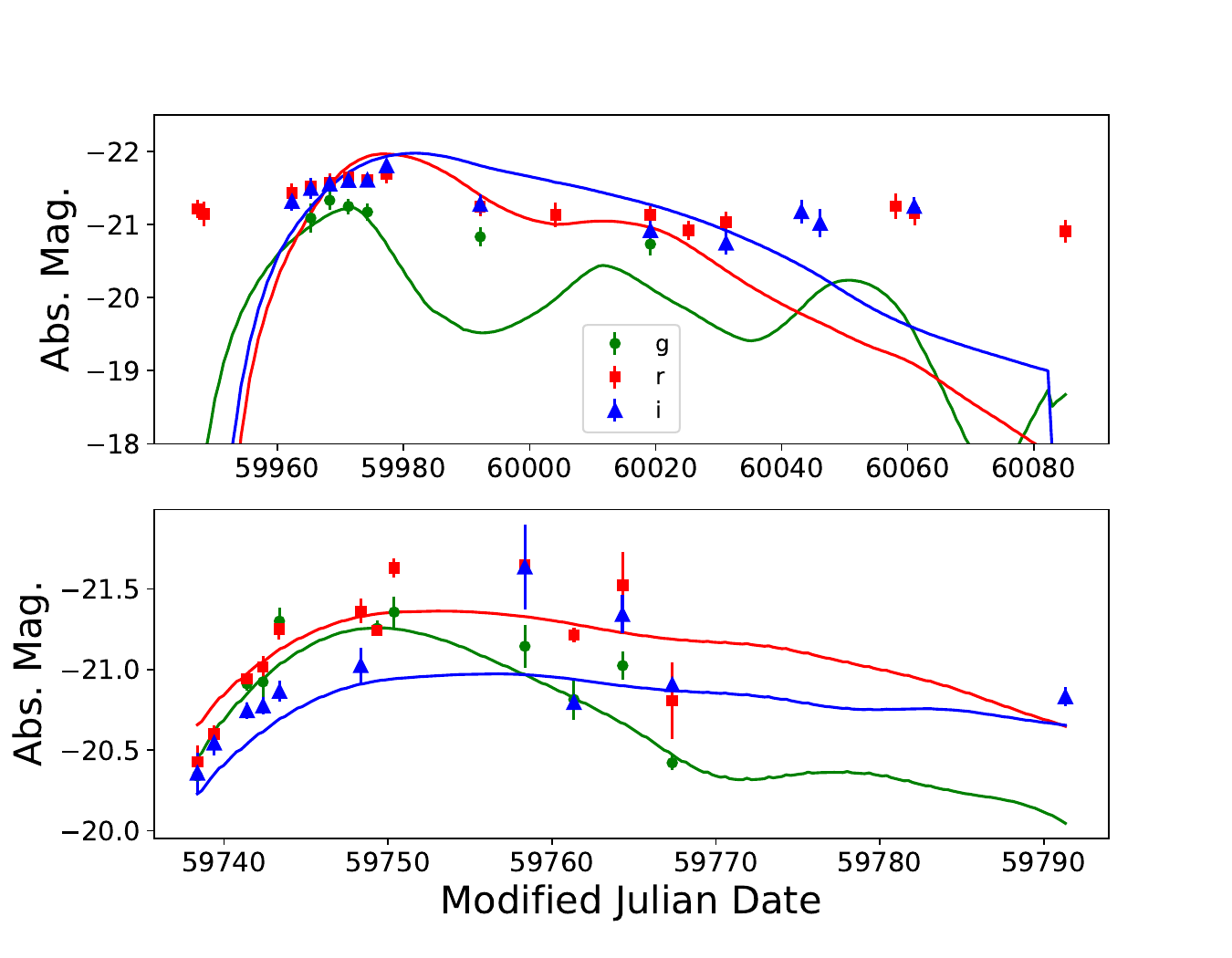}
    \includegraphics[width=.48\textwidth]{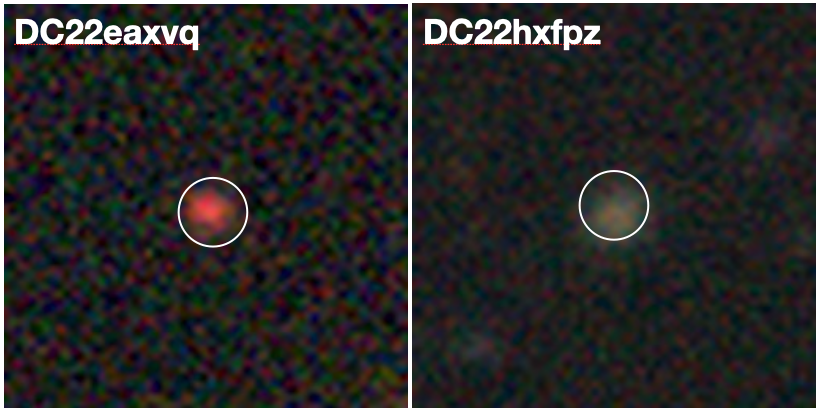}
    \caption{Two most likely lensed SN candidates - DC22eaxvq (top) and DC22hxfpz 
    (middle) - based on their peak brightness from the distance modulus inferred 
    using the photometric redshift as well as visually inspecting the lightcurve 
    to verify SN-like evolution (i.e. an inflection point a few weeks from the 
    first observation). The solid lines are the fits to the doubly imaged SN~Ia model, showing no strong evidence that these candidates are lensed SNe~Ia. At bottom, the Legacy Survey DR9 color images at the 
    locations of the two candidates.}
    \label{fig:lensed_SN}
\end{figure}

From the first method, when using a spectroscopic redshift of the galaxy associated 
to the transient we find 109 objects with an existing redshift. 
We also make the quality cuts applied above for potential SNe, i.e. $>$ 5 nights 
where the transient was detection and a duration $> 10$\,d and $<$ 200\,d. 
The latter cut is made to remove objects that are intrinsically bright and 
long-lived and may have peak luminosities similar to lensed SNe~Ia, e.g. AGNs and TDEs. 
We also make the absolute magnitude cut of $M_g < -20$ mag as defined 
in \citep{Goldstein2017} which leaves us with 2 candidates: DC21cwsvw and DC23ldice. 
DES21cwsvw has only a few points in the $g$ and $r$ filters, making any further 
inference difficult. 
Fitting the lightcurve of DC23ldice, we find that the SN~Ia template fit suggests 
very blue colours which is not expected for lensed SNe~Ia based on the observations 
of recent lensed SNe~Ia \citep{Goobar2017,Goobar2023}, for which the $g-r$ 
colours at peak are $> 0$ mag.  

We analyse the sample with small photo-$z$ errors using the same criteria as the spec-$z$ sample.
From visual inspection we find two candidates: DC22eaxvq and DC22hxfpz that have SN-like 
lightcurves showing both a sharp rise to and decline from peak (Figure~\ref{fig:lensed_SN}). 
DC22eaxvq shows an inflection in the $r$ and $i$ filters and has a peak $g - r$ colour of 0.26 mag. 
One sign that a transient could be a lensed SN is that the SALT2 lightcurve shape 
parameter - $x_1$ - is wider than the typical value of cosmological SNe~Ia, i.e. $ -3 < x_1 < 3$ ($x_1 = 5.95 \pm 0.93$), 
which could be due to the superposition of more than one SN image or because the true redshift 
is higher than the associated redshift.  
From a SALT2 model fit to DC22eaxvq, we infer a large $x_1$ which could suggest that the 
lightcurve is a summation of more than one image with a time-delay. 
Given this lightcurve shape and bright peak luminosity, we test whether a multiply 
imaged SN~Ia model would fit the data better than a single point. 
We use \texttt{sntd} \citep{Pierel2019} to fit a doubly imaged SN~Ia to the data (solid lines in Figure~\ref{fig:lensed_SN}). 
While the doubly imaged SN~Ia model fits better than the single SN~Ia model, the fit only 
converges for fixed magnifications ratios. 
The reduced $\chi^2$ of the fit is $> 10$ suggesting that the multiply imaged SN model does 
not explain the data. 
A spectrum while the transient was live would distinguish whether the transient is a lensed 
SN or an intrinsically bright and wide class. 
A measurement of the source redshift via a host galaxy spectrum will be crucial to verify 
whether the brightness was anomalously high. 
DC22hxfpz shows consistent SALT2 parameters within the distribution of cosmologically 
normal SNe~Ia ($x_1 = 1.8 \pm 0.3$), hence, we do not suspect it is a lensed SN, though in this case as well, 
as verification of the photometric redshift via a host galaxy spectrum will be useful to 
confirm whether the object was brighter than the distribution for normal SNe~Ia. 
We note that if the bright peak magnitude is confirmed for the objects with small photo-$z$ errors, 
they could also be intrinsically luminous transients, e.g. superluminous supernovae \citep{Chen2023}.

\subsection{Candidates Without Static-Sky Matches}\label{ssec:sci_unhosted}

As described in Section~\ref{ssec:obs_xmatch}, 137 candidates were not matched to 
an object in the DESI Legacy Imaging Survey Tractor catalog.
Of these, 86 were only detected on a single night and were probably
asteroids or flaring stars; these were not investigated further in this work.
For the 51 unmatched candidates with multi-epoch photometry, we visually
reviewed the color image stamps from the Legacy Survey.
Of them, 14 had an object within $\sim$10\arcsec\ which \emph{could} have
been associated with the candidate, if it was a transient highly offset
from its host (or if there had been an astrometry error).
One candidate had a small faint object right at its coordinates, DC24kioi, 
the light curve for which appears to be variable (non-transient) and
so is likely an AGN.
The remaining 36 unmatched candidates had no discernible static-sky counterpart.

As a side note, three of 51 unmatched candidates met the initial
light curve constraints to be considered potential SN\,Ia in
Section~\ref{ssec:sci_snia}: DC21dglus, DC21mjrj, and DC23jsner.
Of the three, only DC23jsner met the more strict light curve parameter
cuts to be fit with a SN\,Ia template, and it was also the only one
of the three for which visual review revealed a faint object nearby
(within a few arcseconds) which could be the host galaxy.
None of the other unmatched candidates met the constraints to be
considered potential SNe or fast transients.

\subsubsection{Cross-match to external catalogs}\label{sssec:sci_unhosted_xmatch}

The Vizier service \citep{vizier2000} was used to cross match the
137 unmatched candidates against three catalogs:
SDSS DR16 \citep{2020ApJS..249....3A}, Transiting Exoplanet Survey
Satellite (TESS) Input Catalog v8.2 \citep{2021arXiv210804778P}, and
SIMBAD \citep{2000A&AS..143....9W}.
The cross match for all three catalogs was conducted using a search radius
of 5\arcsec. 

Of the 137 unmatched candidates, 31 were matched with objects in the
SIMBAD catalog.
All but one of these were in the COSMOS field, and mostly matched to
faint, high-redshift galaxies or AGN in the deep catalogs that have been
published for COSMOS (e.g., \citealt{2007ApJS..172...99C}).
The one candidate in ELAIS, DC22ibvrq, was found to have a
bright nearby point-source when the Legacy Survey stamp was reviewed,
and was matched to a QSO published by \citet{2017AJ....153..107T}.

\begin{figure}
\centering
\includegraphics[width=5cm]{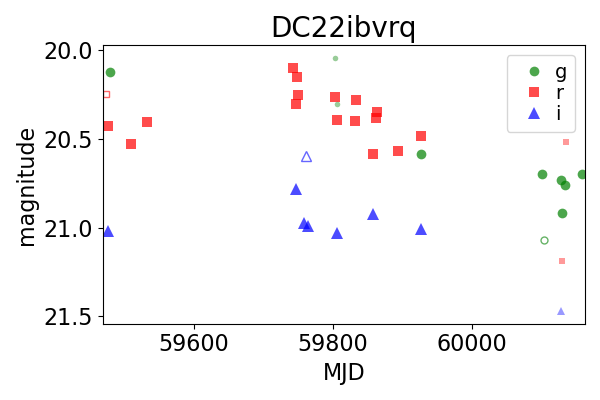} 
\includegraphics[width=3cm]{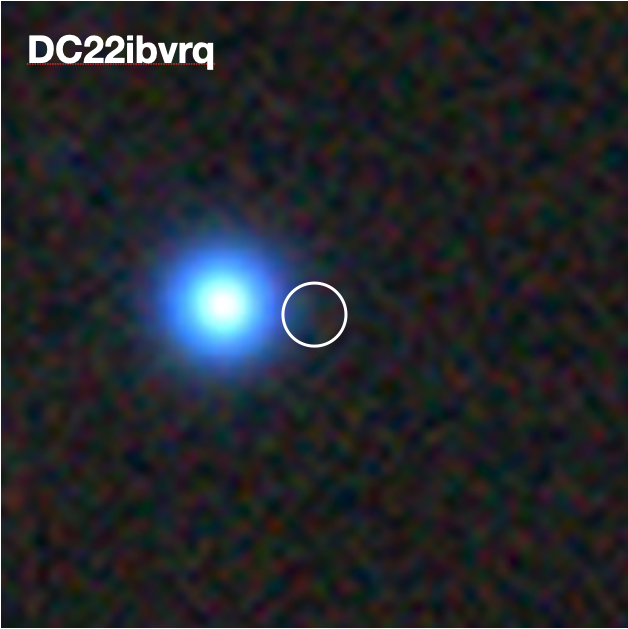} 
\includegraphics[width=5cm]{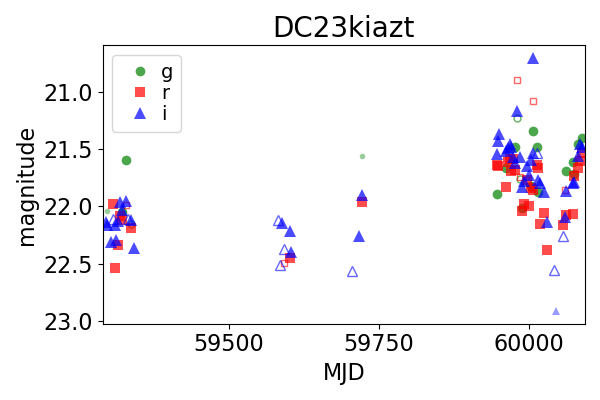} 
\includegraphics[width=3cm]{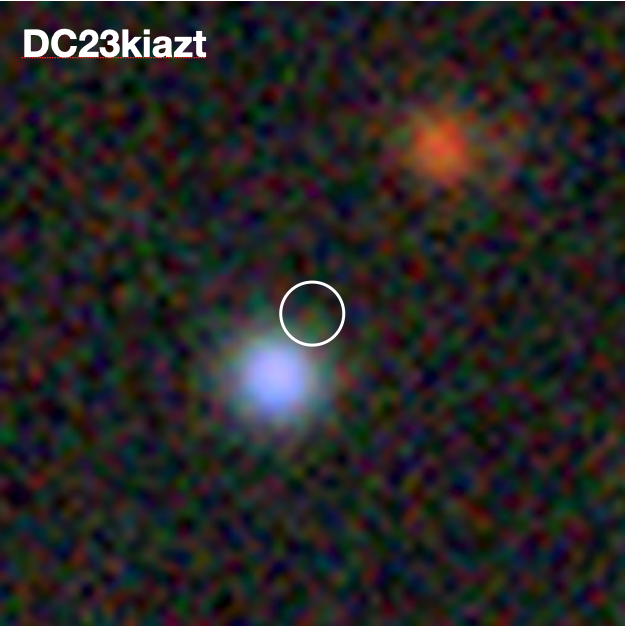} 
\caption{Unmatched candidates DC22ibvrq and DC23kiazt flagged as stars in the 
TESS Input Catalog v8.2, but also cross-matched to a QSO and AGN, respectively.}
\label{fig:DDF_TESS_xmatch}
\end{figure}

From the cross match with the SDSS DR16 catalog, 13 were matched
with objects, all of them were in the COSMOS field as the ELAIS
field was too far south for SDSS.
The matches were mostly with faint, extended SDSS objects, except for
DC23kiazt which was matched with a $g{\sim}$20th magnitude point source.
The light curve of DC23kiazt appears to be variable
(non-transient; Figure~\ref{fig:DDF_TESS_xmatch}),
and it was one of the candidates cross matched with SIMBAD to
an AGN identified by Gaia \citep{2020yCat.1350....0G}.

A cross match with the TESS catalog yielded 3 matches, candidates
DC22ibvrq, DC23jqqgs, and DC23kiazt, all of which were identified
in the TESS catalog as stars.
DC23jqqgs had only a single epoch of photometry, but the light curves
for DC22ibvrq and DC23kiazt are shown in Figure~\ref{fig:DDF_TESS_xmatch}.
DC22ibvrq possesses a TESS luminosity class flag as a dwarf, and
based on Figure~\ref{fig:DDF_TESS_xmatch} it does appear to be
a variable star (or QSO, based on the SIMBAD cross match)
for which our absolute astrometry is slightly off.
As discussed in the paragraphs above, DC23kiazt was also cross matched
to an AGN in SIMBAD.

In summary, despite astrometric errors causing erroneous identifications
of "unmatched" candidates in a few cases, we find that the majority of
the DECam DDF's multi-epoch candidates which were unmatched to a
static-sky counterpart could be associated with high-redshift galaxies
(at least, this is the case in the COSMOS field, where deep legacy
catalogs exist).
A full analysis of what type of phenomena these candidates might
be is left for future work, as here we sought only to characterize
the environments of initially unmatched candidates.

\section{Conclusions and Future Work}\label{sec:conc}

The main message of this analysis is that there is an abundance of scientific
opportunity in the 2,020 candidate light curves which are now publicly
available (see Section~\ref{ssec:obs_dataaccess} for access information).
All are welcome to make use of these data.
The main challenges that we find are that the simple methods used here to identify
\emph{potential} transients and variable stars based on the photometry alone
are not satisfying, and that a comprehensive photometric classification
process is needed.
This work is in progress, but the lack of obvious training sets for 
light curves with the depth and cadence of the survey might be an issue.
Scientific analysis of the DECam DDF would also benefit from having the
pipelines retain objects with \emph{negative} difference-image fluxes,
and produce direct-image magnitudes as well.
This is especially useful for the analysis of non-transient events like AGN and variable stars.
The LSST time-domain data products will include these kinds of measurements.

Of the transients that were classified as AGN or variable stars through cross-matching with DESI spectra, we will extract more detailed light curves without the use of difference imaging in order to get more accurate flux measurements and power spectra.
We will use both aperture and PSF photometry methods for this, as well as a new technique for scene modeling of transient sources that is currently being built into the \texttt{scarlet2} package \citep{scarlet2}.
This computationally improved method for time-domain scene modeling could prove to be more accurate in generating light-curves, which would give us an improved set of reduced data for studying supernova physics and cosmology using DECAT-DDF and, eventually, LSST.

At the time of writing, the DECam DDF was due to restart observations
and alert production during the 2024A semester.
We plan to continue observations with the same three-night cadence (when possible). 
Whenever observing constraints will permit, will plan to also split some of the observations
in the same night into two batches, spaced by a few hours, to allow for 
detection of very fast transient and variable signals. 
For example, as recommended by \citet{2019PASP..131f8002B} for Rubin Observatory's LSST, 
intra-night images separated by $>$1.5 hours are desirable for detecting
evolution in transient light curves within a night.
In this DECam DDF, only 13\% of our individual images were separated by $>$90 minutes,
and as expected we found this insufficient for detecting intra-night evolution.

In addition, a new, more robust pipeline is currently under development, 
which will produce more consistent results, 
based on improved subtraction algorithms and uniform reference images. 
The new pipeline is planned to also include new products 
such as detection of negative transients (e.g., dimming of variables and AGN) 
and nightly stacks.
Forced photometry in all difference- and direct-images,
at the location of all candidates, is left as a 
stretch goal for future work.
Finally, we intend to publish the lightcurves of transients and variables 
along with value-added metadata, such as cross matches to established catalogs, 
redshifts when available (for extra galactic sources), 
mean colors and magnitudes of the objects (for variables) 
and summary statistics such as rise time and maximum magnitude, 
in a database that is searchable and easy to access. 
Until that time, all of the data presented in this
work is available via GitHub
(Section~\ref{ssec:obs_dataaccess})
or by contacting the authors.

\section*{Acknowledgements}

% DECAT

This DECam program for Deep Drilling Fields is a founding member of the DECam Alliance for Transients (DECAT), a logistical solution for a heterogenous group of programs all doing time-domain astronomy on a classically-scheduled telescope.
Within DECAT, multiple DECam programs request that their awarded time be co-scheduled on the Blanco 4m telescope, and then the PIs work together to ensure the targets for all programs are optimally observed, and all program participants share in the observing responsibilities.
We thank all NOIRLab and Blanco staff for their flexibility and support in helping to co-schedule all DECAT programs.

The DECAT nightly observing plans were primarily generated by Gautham Narayan, Dillon Brout, and Armin Rest.
DECAT observers during 2021, 2022, and 2023 were Alejandro Clocchiatti, Alex Gagliano, Alfredo Zenteno, Alice Eltvedt, Amanda Wasserman, Antonella Palmese, Armin Rest, Ben Boyd, Brendan O'Connor, Chris Lidman, Clara Martinez-Vazquez, Colin Burke, Damián Pacheco, Dillon Brout, Divya Mishra, Elana Urbach, Gautham Narayan, Guillermo Damke, Guy Nir, Haille Perkins, Igor Andreoni, Jan Kleyna, Jiawen Yang, Julio Carballo-Bello, Justin Pierel, Kathy Vivas, Keerthi Kunnumkai, Kevin Luhman, Lauren Aldoroty, Maggie Verrico, Matt Grayling, Melissa Graham, Monika Soraisam, Nicole Kane, Noor Ali, Qian Yang, Qinan Wang, Ryan Ridden-Harper, Sammy Sharief, Sasha Brownsberger, Scott Sheppard, Segev BenZvi, Shenming Fu, Tom Shanks, Tomas Cabrera, Tony Chen, Ved Shah, Zach Stone, and Zhefu Yu.

% NERSC & LBNL
% https://www.nersc.gov/users/accounts/user-accounts/acknowledge-nersc/

This research used resources of the National Energy Research Scientific Computing Center, a DOE Office of Science User Facility supported by the Office of Science of the U.S. Department of Energy under Contract No. DE-AC02-05CH11231 using NERSC award ERCAP0024621.
P.E.N. acknowledges support from the DOE under grant DE-AC02-05CH11231, Analytical Modeling for Extreme-Scale Computing Environments.
LBNL is managed by the Regents of the University of California under contract to the U.S. Department of Energy.

% NOIRLab's Astro Data Lab
% https://datalab.noirlab.edu/acknowledgements.php

This research uses services or data provided by the Astro Data Lab at NSF's National Optical-Infrared Astronomy Research Laboratory (NOIRLab).
Based on observations at Cerro Tololo Inter-American Observatory, NSF’s NOIRLab (NOIRLab Prop. IDs 2021A-0113, 2021B-0149, 2022A-724693, and 2022B-762878 under PI M. L. Graham, and 2023A-716082 under PI G. Nir).
NOIRLab is operated by the Association of Universities for Research in Astronomy (AURA), Inc. under a cooperative agreement with the National Science Foundation.

% Legacy surveys requested acknowledgement
% https://www.legacysurvey.org/acknowledgment/

The Legacy Surveys consist of three individual and complementary projects: the Dark Energy Camera Legacy Survey (DECaLS; Proposal ID \#2014B-0404; PIs: David Schlegel and Arjun Dey), the Beijing-Arizona Sky Survey (BASS; NOAO Prop. ID \#2015A-0801; PIs: Zhou Xu and Xiaohui Fan), and the Mayall z-band Legacy Survey (MzLS; Prop. ID \#2016A-0453; PI: Arjun Dey). DECaLS, BASS and MzLS together include data obtained, respectively, at the Blanco telescope, Cerro Tololo Inter-American Observatory, NSF’s NOIRLab; the Bok telescope, Steward Observatory, University of Arizona; and the Mayall telescope, Kitt Peak National Observatory, NOIRLab. Pipeline processing and analyses of the data were supported by NOIRLab and the Lawrence Berkeley National Laboratory (LBNL). The Legacy Surveys project is honored to be permitted to conduct astronomical research on Iolkam Du’ag (Kitt Peak), a mountain with particular significance to the Tohono O’odham Nation.

The Legacy Survey team makes use of data products from the Near-Earth Object Wide-field Infrared Survey Explorer (NEOWISE), which is a project of the Jet Propulsion Laboratory/California Institute of Technology. NEOWISE is funded by the National Aeronautics and Space Administration.

% DECam

This project used data obtained with the Dark Energy Camera (DECam), which was constructed by the Dark Energy Survey (DES) collaboration. Funding for the DES Projects has been provided by the U.S. Department of Energy, the U.S. National Science Foundation, the Ministry of Science and Education of Spain, the Science and Technology Facilities Council of the United Kingdom, the Higher Education Funding Council for England, the National Center for Supercomputing Applications at the University of Illinois at Urbana-Champaign, the Kavli Institute of Cosmological Physics at the University of Chicago, Center for Cosmology and Astro-Particle Physics at the Ohio State University, the Mitchell Institute for Fundamental Physics and Astronomy at Texas A\&M University, Financiadora de Estudos e Projetos, Fundacao Carlos Chagas Filho de Amparo, Financiadora de Estudos e Projetos, Fundacao Carlos Chagas Filho de Amparo a Pesquisa do Estado do Rio de Janeiro, Conselho Nacional de Desenvolvimento Cientifico e Tecnologico and the Ministerio da Ciencia, Tecnologia e Inovacao, the Deutsche Forschungsgemeinschaft and the Collaborating Institutions in the Dark Energy Survey. The Collaborating Institutions are Argonne National Laboratory, the University of California at Santa Cruz, the University of Cambridge, Centro de Investigaciones Energeticas, Medioambientales y Tecnologicas-Madrid, the University of Chicago, University College London, the DES-Brazil Consortium, the University of Edinburgh, the Eidgenossische Technische Hochschule (ETH) Zurich, Fermi National Accelerator Laboratory, the University of Illinois at Urbana-Champaign, the Institut de Ciencies de l’Espai (IEEC/CSIC), the Institut de Fisica d’Altes Energies, Lawrence Berkeley National Laboratory, the Ludwig Maximilians Universitat Munchen and the associated Excellence Cluster Universe, the University of Michigan, NSF’s NOIRLab, the University of Nottingham, the Ohio State University, the University of Pennsylvania, the University of Portsmouth, SLAC National Accelerator Laboratory, Stanford University, the University of Sussex, and Texas A\&M University.

% SIMBAD

This research has made use of the SIMBAD database,
operated at CDS, Strasbourg, France.

% DESI

The Legacy Surveys imaging of the DESI footprint is supported by the Director, Office of Science, Office of High Energy Physics of the U.S. Department of Energy under Contract No. DE-AC02-05CH1123, by the National Energy Research Scientific Computing Center, a DOE Office of Science User Facility under the same contract; and by the U.S. National Science Foundation, Division of Astronomical Sciences under Contract No. AST-0950945 to NOAO.

% BASS

% BASS is a key project of the Telescope Access Program (TAP), which has been funded by the National Astronomical Observatories of China, the Chinese Academy of Sciences (the Strategic Priority Research Program “The Emergence of Cosmological Structures” Grant \# XDB09000000), and the Special Fund for Astronomy from the Ministry of Finance. The BASS is also supported by the External Cooperation Program of Chinese Academy of Sciences (Grant \# 114A11KYSB20160057), and Chinese National Natural Science Foundation (Grant \# 12120101003, \# 11433005).

%%%%%%%%%%%%%%%%%%%%%%%%%%%%%%%%%%%%%%%%%%%%%%%%%%
\section*{Data Availability}
% The inclusion of a Data Availability Statement is a requirement for articles published in MNRAS. Data Availability Statements provide a standardised format for readers to understand the availability of data underlying the research results described in the article. The statement may refer to original data generated in the course of the study or to third-party data analysed in the article. The statement should describe and provide means of access, where possible, by linking to the data or providing the required accession numbers for the relevant databases or DOIs.

See Section~\ref{ssec:obs_dataaccess} for a description and links to
the original images and derived photometry.


\bibliographystyle{mnras}

% if your bibtex file is called example.bib
\bibliography{myrefs} 

%%%%%%%%%%%%%%%%%%%%%%%%%%%%%%%%%%%%%%%%%%%%%%%%%%

%%%%%%%%%%%%%%%%% APPENDICES %%%%%%%%%%%%%%%%%%%%%

% \appendix
% \section{Appendix 1: TBD}
% If an appendix is needed, text goes here.

%%%%%%%%%%%%%%%%%%%%%%%%%%%%%%%%%%%%%%%%%%%%%%%%%%

% Don't change these lines
\bsp	% typesetting comment
\label{lastpage}
\end{document}